\documentclass[aps,prx,noshowpacs,twocolumn,nofootinbib]{revtex4-2}
\usepackage{bm}
\usepackage{graphicx}
\usepackage{mathtools}
\usepackage{color, soul}
\usepackage{graphicx,amsmath,amssymb,amsfonts}

\usepackage{changes}
\usepackage{soul}
\usepackage{hyperref}
\usepackage[11pt]{moresize}

\newcommand{\x}{{\bf x}}

\newcommand{\G}{{\tilde G}}

\renewcommand{\u}{{\bf u}}
\newcommand{\W}{{\bf W}}

\renewcommand{\P}{{\mathcal P}}
\newcommand{\EQ}{\begin{equation}}
\newcommand{\EE}{\end{equation}}
\newcommand{\EQA}{\begin{eqnarray}}
\newcommand{\EEA}{\end{eqnarray}}
\newcommand{\Tr}{{\text{\bf Tr}}}
\renewcommand{\d}{{\text{d}}}
\newcommand{\var}{{\text{var}}}
\renewcommand{\st}{{\text{st}_0}}
\newcommand{\D}{{\mathcal{D}}}
\renewcommand{\L}{{ L}}
\usepackage{xr}
\begin{document}
\title{Optimal evolutionary control for artificial selection on molecular phenotypes}

\author{Armita Nourmohammad}
  \email{Correspondence should be addressed to Armita Nourmohammad: armita@uw.edu}
\affiliation{Department of Physics, University of Washington, 3910 15th Ave Northeast, Seattle, WA 98195}
\affiliation{Max Planck Institute for Dynamics and Self-organization, am Fa\ss berg 17, 37077 G\"ottingen, Germany}
\affiliation{Fred Hutchinson Cancer Research Center, 1100 Fairview Ave N.,  Seattle, WA 98109}
\author{Ceyhun Eksin}%
\affiliation{%
Department of Industrial and Systems Engineering, Texas A\&M University, College Station, TX 77845
}%

\date{\today}

\begin{abstract}
Controlling an evolving population is an important task in modern molecular genetics, including  directed evolution for  improving the activity of molecules and enzymes, in breeding experiments in animals and in plants, and in devising public health strategies to suppress evolving pathogens. An optimal intervention to direct evolution should be designed  by considering its impact over an entire stochastic evolutionary trajectory that follows. As a result, a seemingly suboptimal intervention at a given time can be globally optimal as it can open opportunities for desirable actions in the future. Here, we propose a feedback control formalism to devise globally optimal artificial selection protocol to direct the evolution of molecular phenotypes. We show that artificial selection should be designed to counter evolutionary tradeoffs among multi-variate phenotypes to avoid undesirable outcomes in one phenotype by imposing selection on another. Control by artificial selection is challenged by our ability to predict molecular evolution. We develop an information theoretical framework and show that molecular time-scales for evolution under natural selection can inform how to monitor a population in order to acquire sufficient predictive information for an effective intervention with artificial selection. Our formalism opens a new avenue for devising  artificial selection methods for directed  evolution of molecular functions.
\end{abstract}
\maketitle

\section{Introduction}
The concept of  {feedback control} in molecular evolution was first advocated by {\em A. Wallace} as a way of describing natural selection~\cite{Darwin:1858wo}. Wallace  hypothesized that similar to the centrifugal governor of the steam engine, the action of natural selection is like a {\em controller} that  balances  organismic traits, such that weak feet are often accompanied with powerful wings~\cite{Darwin:1858wo}. Such evolutionary tradeoffs are ubiquitous in natural fitness landscapes. For example, experiments on a protein transport system has shown that the  fitness landscape for the  underlying biochemical network is tuned to exploit optimal control with feedback throughout evolution~\cite{Chakrabarti:2008ct}. However, it remains to be determined  whether these structures are solely reflective of biochemical constraints  or have emerged as incidences of fitness landscapes that could accommodate for  long-term evolutionary survival.

Evolution as a  feedback control is also reflected in the inheritance strategies and phenotypic response of populations  to time-varying environments. A prominent example of such adaptation is observed in bacteria where cells use phenotypic switches to produce slowly replicating bacteria  with tolerance and  persistence  against  antibiotics. Populations use this Lamarckian-type phenotypic response~\cite{lamarck:1809} to hedge their bets against fluctuating environments~\cite{Balaban:2004bq,Paarporn:2018ds}--- an optimal response that can be viewed as an evolutionary feedback control~\cite{Rivoire:2014kt}.

\begin{figure*}
\includegraphics[width =\textwidth]{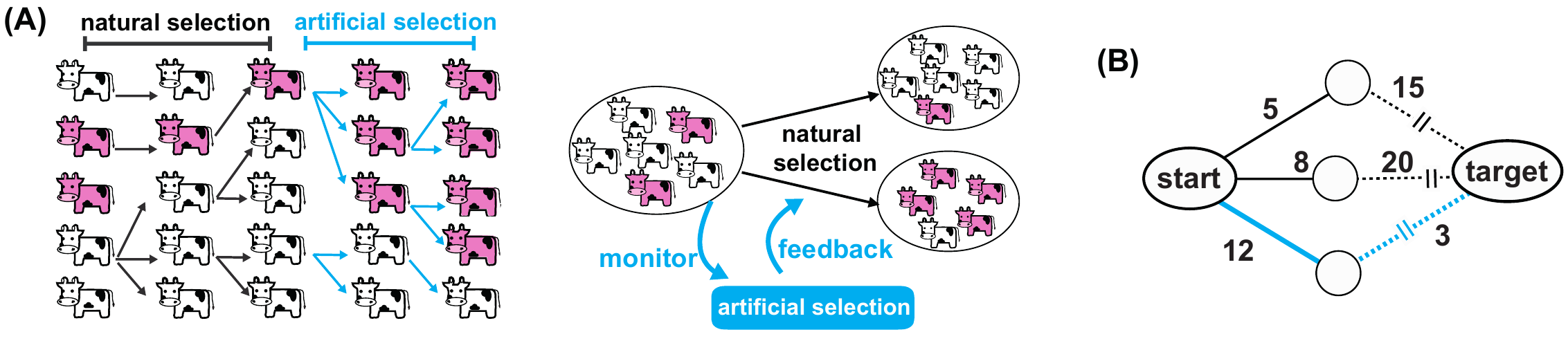}
\caption{{\small {\bf Artificial selection as an optimal stochastic adaptive  control strategy.}
{\bf (A)} Artificial selection is an external intervention to select for a desired trait (i.e., pinkness of cows) in a population, which is otherwise not favored by natural selection. Artificial selection should be viewed as an optimal feedback control, whereby monitoring a stochastically evolving population informs the intervention protocol to optimally bias breeding and reproduction over generations. {\bf (B)} {The schematic graph shows   different paths} with indicated costs for a system to  evolve from a start to a target state. Bellman's principle of optimality states that at each  step an optimal decision is made, assuming that the following  steps are also determined  optimally. Although the first step (full line) of the blue path is more costly compared to the others (dotted lines), its cumulative cost is minimum, and hence, it should be chosen as the optimal path. This decision can be  made best  recursively,  known algorithmically  as dynamic programming. 
 \label{Fig:1}}}
\end{figure*}

Another approach to evolutionary control is through external interventions with artificial selection to direct populations to acquire a desired trait.  Fig.~\ref{Fig:1} demonstrates  artificial selection  with a feedback control to breed ``pink cows", which are otherwise not  favored by natural selection. Such selective breeding has  long been used to domesticate animals or to improve  agricultural yield in crops and   became the basis  for  Mendelian genetics~\cite{Mendel:1862ua}.  

Another important avenue for artificial selection is to characterize intervention protocols against rapidly evolving pathogens, for example to  counter  emergence of drug resistance in bacteria, escape of viral populations from  immune challenge, or progression of evolving cancer tumors~\cite{Fischer:2015dq,Lassig:2020dd}. {As such, control strategies have been  suggested to direct intra-patient somatic evolution of antibody secreting B-cells to elicit  potent broadly neutralizing antibodies against HIV~\cite{Wang:2015em,Shaffer:2016ci,Sprenger:2020jo,Sachdeva:2020gg}. Finding such control strategy involves optimization of the immunization cocktail, and the schedule for immunization, to direct somatic  evolution of antibodies towards an outcome with high potency and breadth.
}

Artificial selection also plays a significant role in improving molecular functions through experimental directed evolution. Importantly, directed evolution in the lab is currently being employed to improve the activity and selectivity of molecules and enzymes~\cite{Eigen:1984,Chen:1993ez,Esvelt:2011cv}, often desirable in industry or for pharmaceutical purposes.   {For example, experimental techniques like {\em morbidostat} have been designed to directly measure the growth rate of evolving microbial populations and accordingly tune the strength of selection induced by antibiotics in order to achieve continuous drug-induced inhibition in an experimental setup~\cite{Toprak:2012ff,Toprak:2013dn}. More recently, control protocols, such as proportional-integral-derivative control~\cite{Ang1999:hh}, have been experimentally implemented in high-throughput continuous directed evolution of proteins to automatically tune artificial selection based on the state of the population and  optimize function~\cite{Wong:2018et,Heins:2019ib,Zhong:2020gs}.  Implementing optimal control in these automated and continuous directed evolution experiments will have significant impact in efficiently synthesizing molecules with desired function.
}

Designing any artificial selection protocol is limited by our ability to predict the outcome of evolution, which is often challenging due to a multitude of  stochastic forces at play, such as mutations, reproductive stochasticity (genetic drift) and  environmental fluctuations~\cite{Nourmohammad:2013in,Lassig:2017hra}. In contrast to strongly divergent evolution at the genetic level, there is growing experimental evidence for convergent predictable evolution at the  phenotypic level~\cite{Toprak:2012ff,Kryazhimskiy:2014kc,BarrosoBatista:2015fna}, including for complex molecular phenotypes like RNA polymerase function~\cite{Tenaillon:2012gd}. We will exploit this evolutionary predictability and focus on designing artificial selection for molecular phenotypes, which are  key links between genotypic information, organismic functions, and evolutionary fitness~\cite{Nourmohammad:2013in}.

Fitness and large-scale organismic traits are often encoded by a number of  co-varying  molecular phenotypes, linked through genetic interactions;  pigmentation patterns on the wings or body of fruit flies are among such multi-dimensional traits, impacted by the expression level of many interacting genes. {A central issue in designing artificial selection for  multi-variate phenotypes is to avoid the undesirable (side)effects of selection, which can arise due to evolutionary tradeoffs, e.g. between thermal stability and function of a protein~\cite{Shoichet:1995ix,Zeldovich:2007hv}.} Evolutionary interventions on multi-variate phenotypes should be designed by assuming their impact over an entire evolutionary trajectory that follows. As a result, a locally optimal action at a given time point may be sub-optimal once considering all the  actions that are required to follow in order to direct the correlated evolution of the phenotypes towards their targets; see Fig.~\ref{Fig:1}B.

Finding a globally optimal protocol to direct a stochastic evolution is a topic of  control theory, known for its impact in engineering, economics  and other fields~\cite{Bertsekas:1995uq}. Here, we introduce a formalism based on  optimal control to devise a population-level artificial selection strategy and drive the stochastic evolution of multi-variate molecular phenotypes towards a desired target. {Importantly, we  develop a framework to quantify how  uncertainty and  lack of evolutionary predictability can limit the  efficacy of such artificial selection. By relating evolutionary predictability with control under artificial selection, we  characterize how to best  monitor a population and acquire a sufficient predictive information in order to optimally intervene with its evolution.}

\section{Results}
\subsection{Model of multi-variate phenotypic evolution under optimal artificial selection}
{ Molecular  phenotypes are often polymorphic due to genetic variation in their encoding sequence within a population. Here, we primarily focus on phenotypes that are encoded by a relatively large number of genetic loci and hence, are approximately normally distributed within a population--- this Gaussian approximation however, can be relaxed as discussed in ref.~\cite{NourMohammad:2016hg}.  In the case of normally distributed $k$-dimensional phenotypes, we  characterize the population's phenotype statistics by the average $\x =[ x_1,x_2,\dots, x_k ]^\top $ and  a symmetric covariance matrix  $K$, where the diagonal elements $K_{ii}(\x)$ indicate the variance of the $i^{th}$ phenotype and the off-diagonal entries $K_{ij}(\x)$ indicate the covariance between different phenotypes. {A non-diagonal covariance matrix reflects existence of  mutational trade-offs between molecular phenotypes, for example between the thermal stability and the catalytic   activity of an enzyme or  function of a protein~\cite{Shoichet:1995ix,Zeldovich:2007hv}, or between the affinity and the breadth (cross-reactivity) of an antibody~\cite{Wedemayer:1997co,frank:2002}.}
 
The primary evolutionary forces that shape the  composition of phenotypes within a population are selection, mutations and genetic drift. Molecular phenotypes are often  encoded in confined genomic regions of about a few 100 bps, and hence, are not strongly impacted by recombination, even in sexually reproducing populations.  The impact of the evolutionary forces on phenotypes can be  directly projected from the evolutionary dynamics in the high-dimensional space of the encoding genotypes~\cite{Nourmohammad:2013ty,Held:2014di}. For Gaussian distributed phenotypes, the change in mean phenotype $d\x$ over a short time interval $\d t$ simplifies to a stochastic process~\cite{Lande:1976ku},
\EQ
d\x = K \cdot \nabla F  \d t +\frac{1}{N} \Sigma \cdot \d \W
\label{eq.NatEvolQuad-main}
\EE
where, $F$ is the adaptive potential and $\nabla F$ is the corresponding adaptive force, reflecting the combined impact of natural selection and mutations during evolution. $\d\W$  is a differential that the reflects the stochastic effect of genetic drift by a multi-dimensional Wiener noise process~\cite{Gardiner:2004tx}. The amplitude of the noise is proportional to  $\Sigma$, which is  the square root of the covariance matrix (i.e., $\Sigma^\top \Sigma\equiv K$), scaled by the effective population size $N$ that adjusts the overall strength of the noise (see Appendix~\ref{popgen} for details). 

The  stochastic evolution of the mean phenotype in eq.~\ref{eq.NatEvolQuad-main} defines an ensemble of evolutionary trajectories. We can characterize the statistics of these evolutionary paths by the dynamics of the underlying conditional probability density $P(\x',t'|\x,t)$ for a population to have  a mean phenotype $\x'$ at time $t'$, given its state $\x$ at an earlier time  $t<t'$.  The dynamics of this probability density  follows a high-dimensional Fokker-Planck equation~\cite{Nourmohammad:2013ty},
{\small
\EQA
\nonumber\frac{\partial}{\partial t} P(\x',t'|\x,t) =
 \left[ \frac{1}{2 } \Tr K \nabla_{\x\x}  - \nabla( K \cdot \nabla F)   \right]P(\x',t'|\x,t)\\
 \label{eq:P_dynamics_main}
\EEA}
\hspace{-1ex} Here, we measured time in units of effective population size ($t \to t/N$), which is the coalescence time in neutrality~\cite{Kingman:1982bk}, and introduced the rescaled adaptive potential $NF \to F$. Moreover, we introduced the compact notation, $\Tr K  \nabla_{\x\x}  \equiv \sum_{ij}K_{ij} \frac{\partial}{\partial x_i}\frac{\partial}{\partial x_j}$.

Similar to the mean phenotype, the covariance matrix $K$ is  a time-dependent variable, impacted by evolution. However,  fluctuations of covariance are much faster compared to the mean phenotype~\cite{Nourmohammad:2013ty,NourMohammad:2016hg}. Moreover, even in the presence of moderately strong selection pressure, the phenotypic covariance depends  only weakly on the strength of selection  and is  primarily determined by the supply of mutations in a population~\cite{Nourmohammad:2013ty,Held:2014di}. Therefore, we  assume that the phenotypic covariance matrix remains approximately constant over time and equal to its stationary ensemble-averaged estimate throughout evolution (Appendix~\ref{popgen}).
 }

\subsection{Artificial selection to optimally  direct evolution}
Natural selection in eq.~\ref{eq.NatEvolQuad-main}  drives populations towards an optimum, which is a function of the organism's environmental and physiological constraints. Artificial selection aims to relax or tighten some of the natural constraints to drive  evolution towards an alternative desired state $\x^*$. In general, we can formulate evolution subject to artificial selection as,
\EQ
d\x = \underbrace{\Big(K \cdot \nabla F + \u(\x,t) \Big)}_{A(\x,t)}\d t+ \Sigma \cdot \d\W\label{eq:langevin_u}
\EE
where $\u(\x,t)$ is a time- and phenotype-dependent vector, which determines the impact of  artificial selection and $A(\x,t)$ is the total  force incurred by natural and artificial selection on the phenotypes.

Our goal is to find an optimal protocol for  artificial selection $\u(\x,t)$ in order to reach the target $\x^*$ by a desired time $t_f$, while minimizing the cost function, 
\EQ{\small
\Omega(\x,\u,t) =  V( \x,t) + \frac{1}{2} \u^\top B \u 
\label{eq:Omega}}\EE
over an entire evolutionary trajectory. Here, $V( \x,t )\equiv V(|\x_t -\x^*|)$ is the cost for deviation of the phenotype state $\x_t$ at time $t$ from the desired target $\x^*$, and $B$ is a matrix that characterizes the cost for imposing artificial selection $\u\equiv\u(\x,t)$ and  intervening with natural evolution of each phenotype.  To solve  the optimal control problem (i.e., to characterize an optimal artificial selection strategy), we  define the {\em cost-to-go} function,
{\small \EQA
\nonumber J(\x,t) = \min_\u \left \langle Q (\x,t_f)+  \int_t^{t_f} ds\left( V( \x_s ) + \frac{1}{2} \u_s^\top B \u_s \right)\right\rangle_{\text{evol.}}\\
\label{eq.cost-to-go}\EEA}
\hspace{-1ex}where the angular brackets $\langle \cdot \rangle$ indicate  expectation over stochastic evolutionary histories from time $t$ until the target time $t_f$. Here,  $Q (\x,t_f)\equiv Q (|\x_{t_f} -\x^*|)$ characterizes the cost of deviation from the target at the end of the evolutionary process $t_f$, which could be chosen to be different from the path cost $V(\x)$. 

\begin{figure*}
\includegraphics[width =\textwidth]{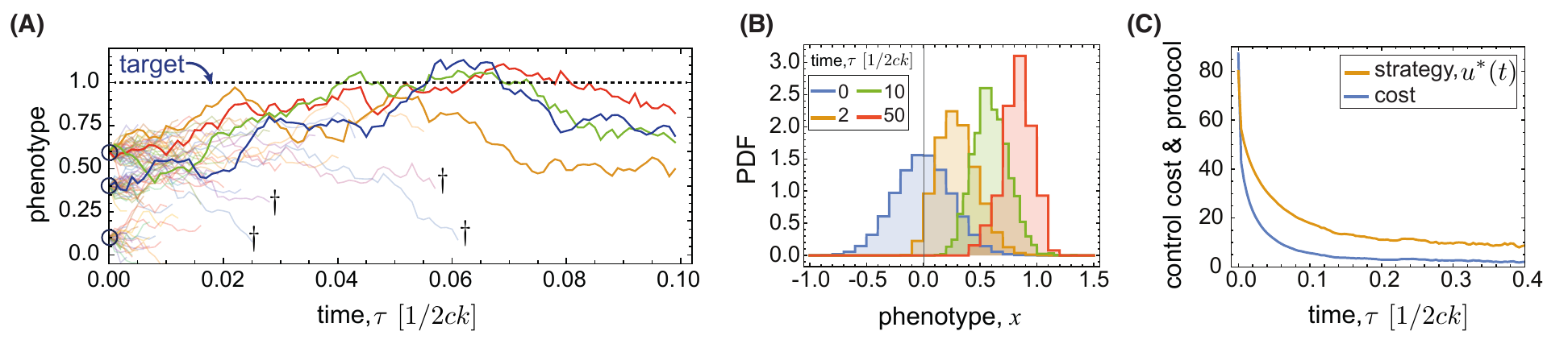}
\caption{{\small
{\bf Artificial selection with stochastic optimal annihilation.}
{\bf (A)} Phenotypic  trajectories  starting from three distinct initial states (open circles) are shown  for evolution under natural selection in a 1D  quadratic adaptive landscape $F(x) = -c x^2$, where $x$ is the phenotype centered around its optimum under natural selection. The trajectories are annihilated ($\dagger$ and low opacity lines) with a rate  proportional to the cost of their deviation from  target throughout evolution (dotted line)  $V(|x-x^*|)/\lambda$ (eq.~\ref{eq:ArtSel}).  At each time point, the survived trajectories characterize the ensemble of evolutionary states for evolution under optimal artificial selection to reach the  target  at $x^*=1$. {Control is designed under the assumption of infinite horizon.  {\bf (B)} The distribution of phenotypes are shown for populations evolving subject to the annihilation protocol in (A). Populations start from an uncontrolled and natural state (blue distribution), and as a result of control annihilation, their distributions move  towards the desired target at  $x^*=1$ (colors). {\bf (C)} The expected control strategy $u^*(t)= -\frac{1}{\lambda} k g (x-x^*)$ (orange) and the cost of control $V(|x-x^*|) + \frac{1}{2} B u^2$ (blue)  are shown as a function of time, as  populations are driven from their natural state towards the target state.}
 Time is measured in  units of the  characteristic time for natural evolution ($1/2k c$). Parameters:  $k=0.4$; $c=1$; $\lambda=0.01$; $g=4$. 
}
\label{Fig:2}}
\end{figure*}

An optimal artificial selection protocol should be designed  by considering its impact over an entire stochastic evolutionary trajectory that follows. As a result, a seemingly suboptimal intervention at a given time can be globally optimal as it can open opportunities for more desirable actions in the future; see schematic Fig.~\ref{Fig:1}B.  An optimal artificial selection protocol at each time point $\mathbf{u}^*(\mathbf{x},t)$ assumes that the selection strategies implemented in the future are also optimal.  This criteria is known as Bellman's ``principle of optimality"~\cite{Bellman:1957tx}, and  allows us to express the optimal control problem in a recursive form,  known as dynamic programming in computer science \cite{Bellman:1957tx} (Appendix~\ref{AppA}). As a result, we can formulate a  dynamical equation for the cost-to-go function, known as Hamilton-Jacobi-Bellman (HJB) equation~\cite{Bellman:1954dx}, 
 
{\small\EQA
\nonumber -\frac{\partial J(\x,t) }{\partial t} 
=\min_\u\left[ \Omega(\x_t ,\u_t)+ A(\x_t)\cdot \nabla J  +  \frac{1}{2} \Tr K  \nabla_{\x\x} J\right]\\\label{eq.HJB}
\EEA}
\hspace{-1ex}with the boundary condition $J(\x,t_f) = Q (\x,t_f)$  at the end of the process (Appendix~\ref{AppA}). Importantly, the HJB equation (\ref{eq.HJB}) indicates that  the cost-to-go  $J(\x,t)$   is a potential function based on which the optimal artificial selection can be evaluated,
\EQ
\u^*(\x,t)= - B^{-1}\cdot \nabla J(\x,t).
\label{eq:Art_fitness}\EE  
In other words,  the cost-to-go function characterizes a time- and phenotype-dependent artificial fitness landscape that determines the  strength of artificial selection $\u^*(\x,t)$.

The solution to the HJB equation (\ref{eq.HJB}) for the cost-to-go function $J(\x,t)$ and the artificial selection $\u^*(\x,t)$ can be complex time- and state-dependent functions, described by non-linear evolutionary operators (Appendix~\ref{AppA}). Here, we consider a class of control problems, known as ``path integral control"~\cite{Kappen:2005bn,Kappen:2005kb,Todorov:2006},  where the cost matrix $B$ for artificial intervention with evolution is inversely proportional to the phenotypic covariance $K$, i.e., $B = \lambda K^{-1}$, where $\lambda$ is a constant that determines the overall cost of artificial selection.  This assumption implies that imposing artificial selection on highly conserved phenotypes is more costly  than on variable phenotypes. This is intuitive as  conserved phenotypes are often essential for viability of an organism and it is best to design a control cost function that limits the access to such  phenotypes through artificial selection. 

The path-integral control assumption results in a significant mathematical simplification for the  dynamics of the  cost-to-go function $J(\x,t)$ and makes the inference of optimal artificial selection more tractable; see Appendices~\ref{AppA},~\ref{AppB}. We can characterize the evolution of the conditional distribution $P_u(\x',t'|\x,t)$  for a population under optimal artificial selection $\u^*(\x,t)$ to be in the phenotypic state $\x'$ at time $t'$, given its state  $\x$ at time $t$ by,
{\small\EQA
&&\nonumber \frac{\partial}{\partial t} P_u (\x',t'|\x,t) \label{rho_dynamics}\\
&&\nonumber= \left[ \frac{1}{2 N} \Tr K  \nabla_{\x\x}  - \nabla( K \nabla F)   - \frac{1}{\lambda} V(\x,t) \right]P_u (\x',t'|\x,t)\\\label{eq:ArtSel}
\EEA}
\hspace{-1ex}with the initial condition $P_u(\x',t|\x,t)= \delta (\x-\x')$ (Appendix~\ref{AppA}). This conditional probability density can be used directly  to compute the cost-to-go function $J(\x,t)$, and consequently the optimal control $\u^*$, as discussed in detail in Appendix~\ref{AppA}.  {Interestingly, the evolution of the optimally controlled conditional distribution $P_u(\x',t'|\x,t)$   resembles the natural evolutionary dynamics (eq.~\ref{eq:P_dynamics_main} with $\u=0$) with an extra annihilation term $ V(\x,t)/\lambda $; see Appendix~\ref{AppA} and ref.~\cite{Kappen:2005kb}. Therefore, artificial selection acts as an {\em importance sampling} protocol over each selection cycle (e.g. each generation) that removes (annihilates) individuals from the population with a rate proportional to their distance  from the evolutionary target $\sim V(|\x_t-\x^*|)/\lambda $; see Fig.~\ref{Fig:2}A. Specifically, at each time point, this protocol generates a phenotypic distribution consistent with the evolutionary process under  optimal artificial selection in eq.~\ref{eq:langevin_u} (Fig.~\ref{Fig:2}B,C), without an explicit knowledge of the  selection protocol $\u^*(\x,t)$; see Appendix~\ref{AppA} and refs.~\cite{Kappen:2005kb}. This result is  highly practical for complex evolutionary processes, for which an analytical description of the optimal control protocol is inaccessible. }

\begin{figure*}
\includegraphics[width =0.95\textwidth]{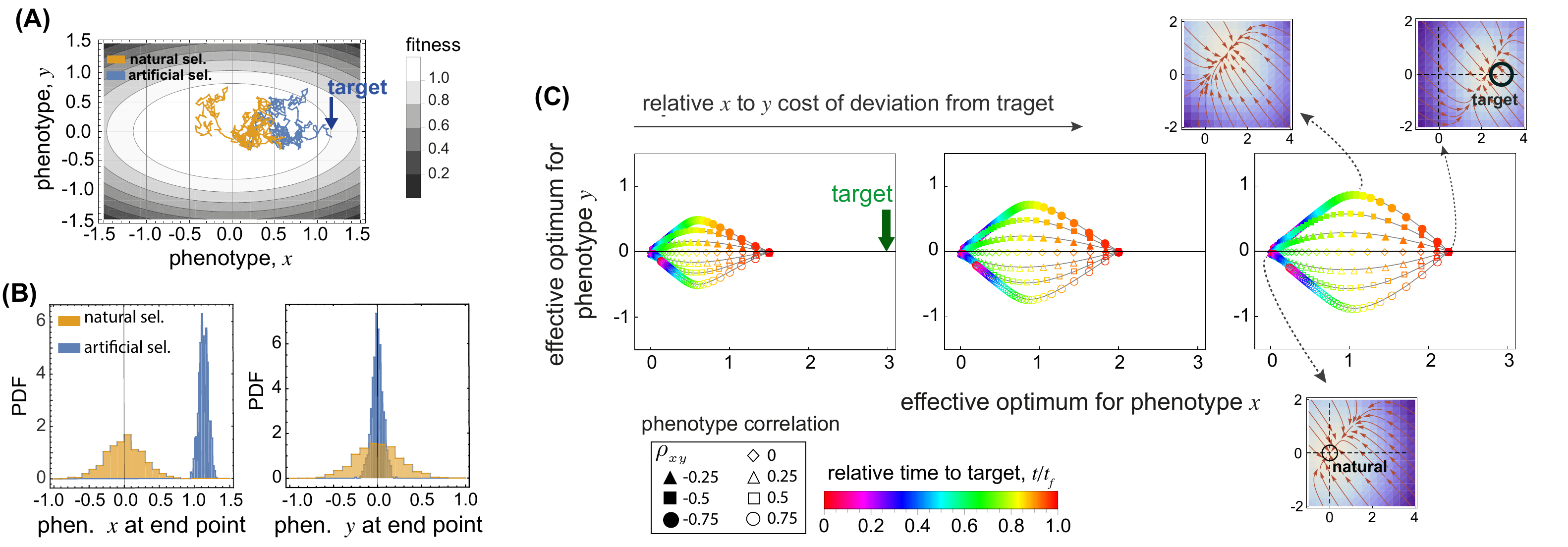}
\caption{{\bf Artificial selection on covarying phenotypes.} {\bf (A)} Trajectories for evolution under natural (orange) and artificial (blue) selection are shown for a 2D phenotype $(x,y)$, in a quadratic landscape. Parameter: $c_x= 2$, $c_y=4$, $c_{xy}=0$; $x^*=1.2$, $y^*=0$; $k_x=0.02$; $k_y=0.05$; $k_{xy}=0$; $g_x=g_y=2$; $\lambda=0.01$. {\bf (B)} The distribution of phenotypes at the end point of an artificial selection protocol  (blue) is compared to the phenotypic  distribution  under natural selection (orange). Evolutionary parameters are  the same as in (A). {\bf (C)} The dynamics of the effective fitness peak is shown over time (colors)  for 2D covarying phenotypes with correlations $\rho_{xy}$ indicated by the shape of the markers. From left to right, panels show increasing end-point cost of deviation from the target along the $x$-phenotype,  $g_x= 1,\,2,\, 3$ , with $g_y=2$. Heatmaps show the effective fitness landscapes associated with a specific fitness peak (indicated by the dotted arrows) for anti-correlated phenotypes  at three different time points. The direction and length of the red arrows in each heatmap indicate the direction and the strength of selection pressure towards the effective optimum.  Parameters: $x^*=3$, $y^*=0$; $c_x=c_y =5$, $c_{xy}=0$;  $k_x=k_y=0.02$;  $\lambda=0.1$.
\label{Fig:4}}
\end{figure*}
{Although cost-to-go function $J(\x,t)$ and optimal control  $\u^*(\x,t)$ are well-known concepts in the field of control theory, their connections to relevant evolutionary measures are far less explored. For evolutionary processes, the scaled cost-to-go-function  $J(\x,t)/\lambda$, can be interpreted as a time- and phenotype-dependent fitness landscape associated with artificial selection $F_{\text{art.}}(\x,t)$; see eq.~\ref{eq:Art_fitness}).} Throughout an artificial selection process, populations evolve in an effective landscape  $\hat F(\x,t) =F(\x) + F_{\text{art.}}(\x,t)$, which is the superposition of the natural fitness landscape $F(\x)$ and the artificial fitness landscape,  $F_{\text{art.}}(\x,t)$. The overall scaling  of the control cost $\lambda$ determines the  impact of artificial selection on evolution relative to  natural selection, and when the control cost is small (i.e., $\lambda \ll 1$), artificial selection can dominate  the course of evolution. 

\subsection{Artificial selection  for multi-variate phenotypes under stabilizing selection}
{Most of our analyses are applicable to general fitness and mutation (i.e., adaptive) landscapes (Appendix~\ref{AppA},~\ref{AppB}). However, we characterize in detail the features of artificial selection to direct evolution on high dimensional quadratic  adaptive landscapes ($F= -\x^\top \cdot C \cdot \x$), in which $C$ is the adaptive pressure and $\x$ is the shifted phenotype vector centered around the optimum under natural selection that the population approaches in stationary state~\cite{Fisher:1930wy}; see Appendix~\ref{popgen}.} In addition, we  assume a  quadratic form for the cost function throughout the evolutionary process, $V (\x,t) = \frac{1}{2}(\x_t-\x^*)^\top G (\x_t-\x^*)$ and also at the end point $ Q (\x,{t_f}) = \frac{1}{2} (\x_{t_f}-\x^*)^\top \tilde G (\x_{t_f}-\x^*)$. 

Characterizing  an artificial selection protocol under such quadratic constraints falls within the class of standard stochastic control problems, known as  linear-quadratic-Gaussian (LQG) control~\cite{Bertsekas:1995uq}. However, we will present our analyses based on the path integral control approach  in eq.~\ref{eq:ArtSel}, which is generalizable beyond LQG and can be applied to arbitrary cost functions and fitness landscapes (see Appendix~\ref{AppB}  and Appendix~\ref{AppC}  for detailed derivation).

Let us imagine that our criteria is to drive evolution towards the optimum $\x^*$ by time $t_f$, which implies that the path cost is zero $G=0$  but the end-point cost is non-zero $\tilde G>0$; see  Appendix~\ref{AppB},~\ref{AppC} for the general scenario including the case with $G>0$. As time approaches the end point, populations transition from evolving in their natural landscape $F(\x)$ to  the artificially induced fitness landscape $F_{\text{art.}}(\x,t_f)$  (Figs.~S1,~\ref{Fig:4}C). Moreover, towards the end point, the  fitness peak and the strength of selection approach their final values, determined by the target and the cost functions in eq.~\ref{eq:Omega}, in an exponentially rapid manner (Appendix~\ref{AppC} and Fig.~S2). Interestingly, at the end point, the optimal artificial selection keeps the population close to the target with a strength,
\EQ
\u^*(\tau\to 0) = -\frac{1}{\lambda} K \tilde G (\x- \x^*)
\label{eq.AsympCon}
\EE
which  resembles  the breeder's equation~\cite{Lush:1937tub} for artificial selection with a heritability factor, $h^2 = K \tilde G/\lambda$; see Appendix~\ref{AppB} for derivations and the general scenario including the case with $G>0$.

One of the main issues in designing  breeding experiments in plants and animals is the  undesirable (side)effects of artificial selection on covarying phenotypes, primarily due to evolutionary tradeoffs~\cite{Lande:1983he} e.g. between sturdiness and flavor of vegetables like tomatoes~\cite{Tieman:2017ky}. Similarly, tradeoffs among covarying molecular phenotypes (e.g. function vs. thermal stability of a protein) could lead to undesirable outcomes for artificial selection at the molecular level. 

To demonstrate the consequences of phenotypic covariation, let us consider a simple example for  artificial selection on two covarying phenotypes ($x,y$); the general solution to this problem in high dimensions is discussed in Appendix~\ref{AppB}. We aim to drive the phenotype $x$ towards the target $x^*>0$ by artificial selection  while keeping the phenotype $y$ at its stationary state value $y^*=0$. An optimal artificial selection protocol  defines an effective two-dimensional quadratic fitness landscape that  biases the  evolutionary trajectories towards the target state (Fig.~\ref{Fig:4}A). As a result,  the phenotype distributions at the end of the process become significantly distinct from the expectation under natural selection, and remain strongly restricted around their  target values; Fig.~\ref{Fig:4}B.

The peak of this fitness landscape (i.e., the effective optimum) changes from the natural state $(0,0)$ to the target state $(x^*,y^*)$ by the end of the  selection process; Fig.~\ref{Fig:4}C and Fig.~S3.  The fitness peak moves monotonically along the $x$-phenotype from the natural optimum $0$ towards the target  $x^*$, irrespective of the correlation $\rho_{xy}$ between the two phenotypes; Fig.~\ref{Fig:4}C. However, the dynamics of the  fitness peak along the $y$-phenotype is generally  non-monotonic and strongly dependent on the phenotypic correlation $\rho_{xy}$.  An increase in $x$  drives the positively (negatively) correlated $y$ towards higher (lower)  values. Therefore,  in the beginning of the process,  the optimal artificial selection protocol sets the fitness peak for the $y$-phenotype at an opposite direction to  counter-balance the effect of  evolutionary forces due to phenotypic covariation. As the end-point approaches, artificial selection becomes significantly strong with an effective fitness optima set at the target  for each phenotype $x^*$ and $y^*$ (eq.~\ref{eq.AsympCon}). Therefore, the optimum $y$-value should return to its target state ($y^*=0$), resulting a non-monotonic behavior in the dynamics of the fitness peak along the $y$-phenotype; see Fig. ~\ref{Fig:4}C. Moreover, the strength of selection also transitions over time and becomes stronger towards the target phenotypes at the end point (heatmaps in Fig.~\ref{Fig:4}C and Fig.~S4).\\

{The optimal  artificial  selection protocol  in Fig.~\ref{Fig:4} is sensitive to the structure of the phenotypic covariance matrix $K$ (eq.~\ref{eq.AsympCon} and Appendix~\ref{AppAltSel}). Importantly, disregarding the phenotypic covariance in designing a control protocol would result in an increase in the associated cost and failure to reach the desired phenotype targets, as the controller misjudges the response of the covarying phenotypes to the designed interventions (Fig.~S2 and Appendix~\ref{AppAltSel}). 
 
The optimal artificial selection protocol requires an accurate representation of the evolutionary dynamics which may not be available in certain scenarios. In such settings, we can devise a na\"{i}ve but an intuitive control strategy that is informed by the structure of the optimal control. For instance, optimal controller with quadratic cost on a quadratic adaptive landscape (LQG) is proportional to the difference between the phenotypic state and the target ($\mathbf{x}_t-\mathbf{x}^*$) with a pre-factor (i.e., a relative strength) that exponentially increases as the  time approaches the end-point (Appendix~\ref{AppC} and Fig.~S2). Using the general structure of this optimal strategy, we can devise proportional controllers with exponentially growing weights even when the accurate model for the underlying dynamics is not available (Appendix~\ref{AppAltSel}).
Importantly, the protocol inspired by the optimal control outperforms a completely na\"ive proportional controller that only intervenes with the evolutionary dynamics as the deviation of  the population  form the target  passes a threshold but does not tune its relative strength as time approaches the end point; Fig.~S5 and Appendix~\ref{AppAltSel}. However, it should be noted that the parameters  of the exponential-proportional controller (e.g. the exponential weights) should be tuned  to achieve the desired target states which may be time-consuming in practice. Moreover, tuning these weights according to certain objectives, e.g., achieving the desired phenotype traits, may have undesirable consequences such as high variability or large cost of control (Fig.~S5 and Appendix~\ref{AppAltSel}).
 }
\begin{figure*}
\includegraphics[width =\textwidth]{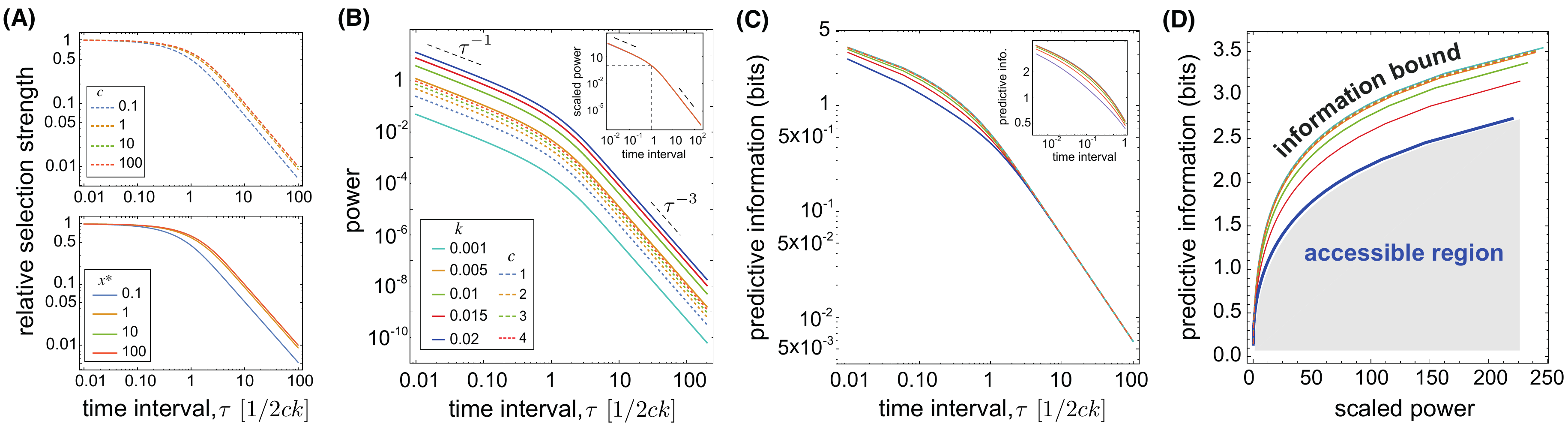}
\caption{
{\bf Artificial selection with limited information.} {\bf (A)} Relative  strength of artificial selection $\alpha_\tau/\alpha_0$ (eq.~\ref{eq.alpha}) is shown as a function of the time interval for monitoring and intervention $\tau$, measured in units of the characteristic time for evolution under natural selection ($1/2kc$).  The selection ratio is shown for various strengths of natural selection $c$ (top; with $x^* = 1$) and for various targets of artificial selection $x^*$ (bottom; with $c=1$). {\bf (B)} Power (eq.~\ref{eq.power}) is shown as a function of the time interval $\tau$  for a  range of  parameters for the phenotypic diversity $k$ (full line) and the strength of natural selection $c$ (dotted line). The insert shows a collapse plot for power scaled with the expectation at the characteristic time for natural selection $\text{Power}(\tau)/\text{Power}(\tau=1)$. {\bf (C)} Predictive mutual information $\mathcal{I}(\tau)$ (eq.~\ref{eq.PredMutInfo}) is shown to  decay with the time interval $\tau$ for  a wide range of  parameters ($k,\, c$). Insert zooms into a narrower range for the time interval $\tau<1$.  {\bf (D)} Predictive information (eq.~\ref{eq.PredMutInfo}) is shown as a function of the scaled power for optimized artificial selection  for a range of $\tau$ values (eq.~\ref{eq.power}). Each curve sets an information bound for artificial selection for a given set of evolutionary parameters ($k,\, c$). A non-optimal  selection intervention should lie below the information curve, shown by the gray shaded area as the accessible region associated with the dark blue curve. Color code in (C,D) is similar to (B). Other parameters: $\lambda=0.6$; $x^*=3$; $g=2$. 
\label{Fig:5}}
\end{figure*}
\subsection{Artificial selection with intermittent monitoring}
Imposing  artificial selection based on continuous feedback from the state of the population  (Fig.~\ref{Fig:1}) requires  complete monitoring and the possibility of continuous  evolutionary intervention--- a criteria  that  is often not met in real conditions. In general, discrete selection protocols based on a limited information  can be inefficient for devising evolutionary feedback control ~\cite{Rivoire:2011ue,Fischer:2015dq}.  {Here, we develop a principled approach to characterize the limits of discrete optimal interventions based on the  evolutionary response of the population to artificial selection.} We consider a simple scenario where  in a stationary state we aim to keep a population at the target phenotype $\x^*$, using discrete monitoring and  interventions  at  time points ($i= 1,\dots, M$) with a time separation $\tau\equiv t_{i+1}-t_i $. We define a   stationary cost-to-go function, 

{\small \EQA
\nonumber&&J(\x,t_m;\tau) \\
\nonumber &&=\min_{\u}  \lim_{M\to \infty} \frac{1}{(M-m)\tau} \left\langle \sum_{i=m}^M \u_i^\top B \u_i + \int_{t_i}^{t_M} V(\x_t)\d t \right\rangle_{\text{evol.}}\\
\label{eq:Stat.J_main}
\EEA}
\hspace{-1ex}where the division by the total time $M\tau$ assures that the cost-to-go remains finite. To further simplify, we  only consider one dimensional phenotype $x$ with intra-population variance $k$, the cost of   deviation $V(x) = g(x-x^*)^2/2$ from target $x^*$, and the cost of intervention $\beta u^2/2$  with  artificial selection $u$. However, our analyses can be easily generalized to multi-variate phenotypes.

In the stationary state and in the regime of small perturbations ($g k/\lambda \ll1$), the optimal artificial selection protocol $u^*$ should be a variant of the case with full information with a strength of selection $\alpha_\tau$ dependent on the time window $\tau$, $u^*_\tau=-k \alpha_\tau (x-x^*)$; see Appendix~\ref{AppD}. We can characterize the optimal strength of artificial selection $\alpha_\tau$ by minimizing the cost-to-go function in eq.~\ref{eq:Stat.J_main}, 
\EQA
\nonumber\alpha_\tau =\alpha_0\left[ \frac{  (1- e^{-2  \tau} ) + 8 c ( x^*)^2(1- e^{- \tau}) }{  2\tau (1 + 4 c (x^*)^2)}\right] + \mathcal{O}\left( (k\gamma/\lambda)^2\right)\\
\label{eq.alpha}\EEA
where $\alpha_0= g/\lambda$ is the optimal selection strength under continuous monitoring. Here,  time $\tau$ is measured in units of the characteristic time for evolution under natural selection, i.e., $(2kc)^{-1}$. 

The  partially informed  artificial selection $\alpha_\tau$ depends on most of the evolutionary parameters similar to selection with complete monitoring $\alpha_0$. Importantly, the ratio $\alpha_\tau/\alpha_0$,  depends only weakly  on the  strength of natural selection $c$   (Fig.~\ref{Fig:5}A; top) and the target for artificial selection $x^*$ (Fig.~\ref{Fig:5}A; bottom) and it is insensitive to the phenotypic diversity $k$ and the parameter $\lambda$ (eq.~\ref{eq.alpha}). 

However, the optimal artificial selection $\alpha_\tau$ strongly depends on the  time interval $\tau$ and it decays as the time interval $\tau$ broadens  (Fig.~\ref{Fig:5}A). This decay is linear and relatively slow up to the characteristic  time for evolution under natural selection $ (2kc)^{-1}$. This is the time scale over which an intermittent artificial selection can  still  contain the population around the desired target $x^*$. If interventions are  separated further in time  (i.e., $\tau \gg 1$), the optimal selection strength decays rapidly as  $\sim\tau^{-1}$. Imposing a strong artificial selection in this regime is highly  ineffective as  populations can easily drift away from the target and towards their natural state within each time interval, and any artificial selection would only contribute to the intervention cost $\sim u^2$ without offering any benefits.

\subsection{Information cost for artificial selection}
Artificial selection  is an evolutionary drive that shifts the equilibrium state of the population under natural selection to a new state around the target. As a result, the phenotypic histories  $\x_{t_0,\dots t_f}$ over the period of $(t_0,\dots,t_f)$ are statistically distinct  for evolution under natural and artificial selection (Fig.~\ref{Fig:4}A). This deviation can be quantified by the Kullback-Leibler distance $D_{KL} (\P_u (\x)|| \P(\x))$ between the distribution of  histories under artificial selection $\P_u(\x) \equiv \P_u(\x_{t_0,\dots t_f})$ and under natural selection $\P(\x)$. In the stationary state, the Kullback-Leibler distance quantifies the impact of artificial selection on evolution and can  be  interpreted as the amount of work $W_{t_0}^{t_f}(\u)$ done by external forces~\cite{Landau:1958up} (i.e., the artificial selection) to shift the population from its natural equilibrium to the artificial equilibrium,

 \EQA
\nonumber W_{t_0}^{t_f}(\u)&=&D_{KL}(\P_u (\x)|| \P(\x)) \\
&=&\int d\x_{t_0}^{t_f}\, \P_u (\x)\log \left [\frac{\P_u (\x)}{\P (\x)}\right]  \label{eq.work}\EEA

The cumulative work is related to the cost of artificial selection, and for the case of path integral control, it is equivalent to the  cumulative  cost of control {\small $W_{t_0}^{t_f}(\u)=\langle\frac{1}{2} \int  \u^\top K^{-1} \u\, \d t \rangle =\frac{1}{2\lambda} \langle \int  \u^\top B \u \,\d t \rangle$}, where the angular brackets $\langle\cdot \rangle$ denote expectation over the ensemble of evolutionary histories under artificial selection; see refs.~\cite{Kappen:2016he,Theodorou:2012gx} and Appendix~\ref{AppE}. The power (i.e., work per unit time), associated with intermittent artificial selection, can be expressed as the amount of work  per time interval $\tau$,
{\small \EQA
 \text{Power}(\tau)= \lim_{M\to \infty}\frac{1}{M\tau} \sum_{i=1}^M W(t_i) =    \frac{1}{2\tau} \langle \u_\tau^\top K^{-1} \u_\tau \rangle
\label{eq.power}
 \EEA}

The expected work, and hence the power, depend on the time interval $\tau$ through various factors. Work depends quadratically on the strength of artificial selection $\alpha_\tau$ and on the  expected population's deviation from the target $\langle (x_\tau-x^*)^2\rangle$. On the one hand, the optimal strength of artificial selection $\alpha_\tau$ decays with increasing the time interval; see Fig.~\ref{Fig:5}A  and  eq.~\ref{eq.alpha}. On the other hand, as the time interval broadens,  populations deviate from the target  towards their natural state,  resulting in an increase in the expected work by artificial selection.  Moreover, since  interventions are  done once per cycle,  the  power  has an overall  scaling with the  inverse of the cycle interval $\sim \tau^{-1}$. These factors together result in a reduction of the expected power associated with artificial selection as the time interval widens; see Fig.~\ref{Fig:5}B. 
 
Power depends strongly on the parameters of natural evolution including the strength of natural selection ($c$) and the phenotypic diversity within a population ($k$); Fig.~\ref{Fig:5}B. This is due to the fact that steering evolution under strong natural selection (i.e., with large $k$ and $c$) is more difficult and would require a larger power by artificial selection. However, the dependence of power on the evolutionary parameters ($k$, $c$) remain approximately separated from its dependence on the time interval $\tau$. Thus, power rescaled by its expectation at the characteristic time $\tau=1$ shows a universal time-decaying behavior, independent of the  evolutionary parameters (Fig.~\ref{Fig:5}B).

\subsection{Predictive information as a limit for efficient artificial selection}
Artificial selection can only be effective to the extent that an intervention is  predictive of the state of the population in the future. The mutual information  between  artificial selection and the future state of the population quantifies the amount of predictive information~\cite{Bialek:1999bc} by artificial selection, or alternatively, the memory of the population from the selection intervention. 
We characterize the predictive  information $\mathcal I_\tau$  as a time-averaged mutual information $I(x_{t},x_0)$ between an intervention (at time $t=0$) and the state of the population at a later time $t,\, (0<t<\tau)$, during each intervention cycle in the stationary state, 
\EQA
\nonumber \mathcal{I}_\tau &=& \frac{1}{\tau} \int_0^\tau \d t\, I (x_{t},x_0) \\
\nonumber&= &\frac{1}{\tau} \int   \d t \int dx_0 dx_t P(x_t, x_0) \log \left[ \frac{P(x_t| x_0)}{P(x_t)}\right]\\
\label{eq.PredMutInfo}
\EEA
The predictive mutual information monotonically decreases as the time interval $\tau$ increases and the population evolves away  from the selection target; see Fig.~\ref{Fig:5}C.

Predictive information in eq.~\ref{eq.PredMutInfo} quantifies the impact of artificial selection on the future of a population. The information theoretical measure  of power  in eq.~\ref{eq.power} on the other hand, quantifies how the optimal artificial selection protocol distinguishes a population's past evolutionary history from the expectation under natural selection. The efficacy of any intervention (i.e., power) is tightly bounded by the impact it may have on the state of the population in the future (i.e., predictive information); see Fig.~\ref{Fig:5}D. Any non-optimal artificial selection effort should lie below this bound and within the accessible region of the information-power plane (Fig.~\ref{Fig:5}D). 

Phenotypic diversity $k$ characterizes the rate at which a population evolves away from the target and towards its natural state during an intervention interval (eq.~\ref{eq.NatEvolQuad-main}). As a result, the information bound for artificial selection is tighter in more diverse populations, which can rapidly evolve away from the target and towards their natural state during each time interval $\tau$. 

As interventions become more frequent,  predictive mutual information increases but  more slowly than the  amount of power necessary to induce an effective artificial selection (Fig.~\ref{Fig:5}D). Importantly, the gain in predictive information becomes much less efficient for time intervals shorter than the characteristic time of natural selection ($\tau\ll 1$). 

{We postulate that trading power with information provides a guideline for scheduling of control interventions of stochastic processes in general, and for evolutionary control, in particular. The characteristic time for evolution under natural selection is a powerful gauge for scheduling the interventions. Importantly, setting the time interval within the range of the characteristic evolutionary time $\tau \sim 1$ could provide a sufficient power-to-information ratio for an optimal artificial selection protocol. However, as information becomes less predictive or the inferred selection protocol becomes sub-optimal, it would be necessary to monitor and intervene more frequently.

Predictive information quantifies how the state of the system in the past is informative of its future, whereas the control power measures the cost associated with an intervention due to its impact on the future state of the system. The connection between predictive information and control  in the context of directed evolution relates the past and the future  of an evolutionary process, subject to external interventions. Indeed, predictive information sets the limit for an effective control in general stochastic processes, but the interpretation of power and predictive information would be specific to the problem in hand.
}

\section{Discussion}
An optimal  intervention should be designed by considering its impact over an entire evolutionary trajectory that follows. Here, we infer an artificial selection strategy as an optimal control with feedback to drive multi-variate molecular phenotypes towards a desired target. This selection protocol is optimal over the defined time course and may seem sub-optimal on short time-scales as it  drives one phenotype  away from its target while driving another towards the target to overcome tradeoffs (Fig.~\ref{Fig:4}C). Monitoring  and feedback from the state of a population are key for imposing an effective artificial selection strategy. 
We show that the schedule for monitoring  should be informed by the molecular time-scales of evolution under natural selection, which set the rate at which a population loses memory of artificial interventions by evolving towards its favorable state under natural selection. 

Being able to control evolution could have significant applications in designing novel molecular functions or in suppressing the emergence of undesirable resistant strains of pathogens or cancers. {Interventions that select for desired phenotypes have become possible in molecular engineering~\cite{Cobb:2013eh,Hess:2017ew,Zhong:2020gs}, in targeted immune-based therapies against evolving pathogens~\cite{Naran:2018ez}, and in immunogen design for optimal vaccination protocols against rapidly evolving viruses like HIV~\cite{Wang:2015em,Shaffer:2016ci,Sprenger:2020jo,Sachdeva:2020gg}. 

One class of experiments for artificial selection uses targeted mutations that are inferred to be beneficial, using machine learning techniques to  characterize genotype-phenotype maps~\cite{Sinai:2020vf}. Another class of experiments relies on implementing feedback control to tune artificial selection during continuous evolution of molecules and proteins, to direct  them towards a desired target~\cite{Toprak:2012ff,Toprak:2013dn,Wong:2018et,Heins:2019ib,Zhong:2020gs}. For example, implementing proportional-integral-derivative (PID) control~\cite{Ang1999:hh}, which is known for its role in cruise control during driving, has shown significant improvements for molecular optimization by directed continuous evolution~\cite{Zhong:2020gs}.  PID control is simple to implement in practice and it is relatively robust to errors, as its proportional term corrects for spontaneous error, the integrator reduces the impact of  long run error, and derivative term would suppress overshooting by anticipating the impact of control~\cite{Bertsekas:1995uq}.  Still, PID is far from optimal and requires fine-tuning of control parameters for the system, which is  often tedious and subject to uncertainty. Therefore, implementing a more principled optimal control approaches for molecular evolution, such the path integral control introduced here, can further optimize the continuous directed evolution  experiments.
}

The efficacy of these actions are limited by our ability to monitor and predict the evolutionary dynamics in response to interventions. Evolution is shaped by a multitude of stochastic effects, including the stochasticity in the rise of novel beneficial mutations and  fluctuations in the environment, which at best, raise skepticism about predicting evolution~\cite{Nourmohammad:2013in,Lassig:2017hra}. However, evolutionary predictability is not an absolute concept and it depends strongly on the time window and the molecular features that we are interested in. For example, despite a rapid evolutionary turnover in the influenza virus, a number of studies have successfully forecasted the dominant circulating strains for a one year period~\cite{Luksza:2014hj,Neher:2014eu}. Similarly, phenotypic convergence across parallel evolving populations has been reported as an evidence for phenotypic predictability, despite a wide genotypic divergence~\cite{Toprak:2012ff,Tenaillon:2012gd,Kryazhimskiy:2014kc,BarrosoBatista:2015fna}. Therefore, to exploit the evolutionary predictability  for the purpose of control, it is  essential to identify the relevant  degrees of freedom (e.g.,  phenotypes vs. genotypes) that confer predictive information and to characterize the evolutionary time scales over which  our observations from a population can inform our interventions to drive the future evolution. 

We focus on modeling control of molecular phenotypes. Phenotypic diversity  within a population  provides standing variation that selection can act upon. To allow for a continuous impact of artificial selection over many generations, we have limited our analyses to a regime of moderate time- and phenotype-dependent artificial selection  to sustain the phenotypic variability in a population.  However, it would be interesting to allow for stronger artificial selection to significantly impact the standing variation and the structure of the phenotypic covariance within a population over time. Indeed, allowing a population to rapidly collapse  as it approaches a  desired target is a common strategy in evolutionary optimization algorithms~\cite{Otwinowski:2020kd}--- a strategy that could accelerate the  design of new functions with directed molecular evolution. 

In this work, we  assume a stochastic model for evolution of multi-variate molecular phenotypes, which has been powerful in describing a  range biological systems, including the evolution of gene expression levels~\cite{Nourmohammad:2017is}. Indeed, optimal control protocols are often designed by assuming a specific  model for the underlying dynamics.  However, in most biological systems, we lack a knowledge of the details and the  relevant parameters of the underlying evolutionary process. {Optimal control strategies can  inform ad-hoc (albeit suboptimal) control approaches   to drive evolution towards the desired target (Fig.~S5 and Appendix~\ref{AppAltSel}).  In addition, if one can at least approximately design a control scenario that satisfies the criteria for path integral control, an effective artificial selection protocol could be achieved through annihilation of evolutionary trajectories with a rate proportional to the cost of their deviation from the desired target (Fig.~\ref{Fig:2}), and without a detailed knowledge of the underlying dynamics. Nonetheless, the ultimate goal is  to simultaneously infer an effective  evolutionary model  based on the accumulating observations and to design an optimal intervention to control the future evolution--- an involved optimization problem known as dual adaptive control~\cite{Wittenmark:1995vc}. }

\section*{Acknowledgements}
We are thankful to Bert Kappen, Jakub Otwinowski, and Olivier Rivoire for discussions and providing valuable feedback. This work has been supported by Royalty Research fund from the University of Washington (AN),  the DFG grant (SFB1310) for Predictability in Evolution (AN, CE), the MPRG funding through the Max Planck Society (AN), and the National Science Foundation grant NSF ECCS-1953694 (CE). 
\clearpage{}
\newpage{}

\onecolumngrid
\appendix
{
\section{Evolution of multi-variate molecular phenotypes}
\label{popgen}
In the case of normally distributed $k$-dimensional phenotypes, we  characterize the population's phenotype statistics by the average $\x =[ x_1,x_2,\dots, x_k ]^\top $ and  a symmetric covariance matrix  $K$, where the diagonal elements $K_{ii}(\x)$ indicate the variance of the $i^{th}$ phenotype and the off-diagonal entries $K_{ij}(\x)$ indicate the covariance between different phenotypes. 
To model the evolution of such multi-variate phenotype, we consider the three primary evolutionary forces: natural selection, mutations and genetic drift. The effect of selection on the mean phenotype is proportional to the covariance  between fitness and phenotype within a population~\cite{Price:1970ez}.  For Gaussian distributed phenotypes, the change in mean phenotype $d\x$ over a short time interval $\d t$ simplifies to a stochastic process~\cite{Lande:1976ku},
\EQ
d\x = (K \cdot \nabla F + \nabla M) \d t +\frac{1}{N} \Sigma \cdot \d\W
\EE
where, $F$ and $M$ are fitness and mutation potentials, respectively. 
The gradient functions (denoted by  $\nabla F$ and $\nabla M$) determine the forces acting on the phenotypes by selection and mutation, respectively~\cite{Nourmohammad:2013ty}. $\d\W$  is a differential that the reflects the stochastic effect of genetic drift by a multi-dimensional Wiener noise process~\cite{Gardiner:2004tx}. The amplitude of the noise is proportional to  $\Sigma$, which is  the square root of the covariance matrix (i.e., $\Sigma^\top \Sigma\equiv K$), scaled by the effective population size $N$ that adjusts the overall  strength of the noise. In other words, the fluctuations of the mean phenotype across realizations of an evolutionary process is proportional to the intra-population variance $K$ and inversely scales with the effective population size (i.e., the sample size) $N$. 

The fitness potential $F$ can be simply approximated by the mean fitness of a population~\cite{Nourmohammad:2013ty}. The mutational potential $M$ however, can depend on the underlying genotype-phenotype map. To characterize $M$, let us assume a general map from a genotypic sequence with $\ell$ loci $\vec \sigma = (\sigma_1,\sigma_2,\dots,\sigma_\ell)$ and the encoded $k$-dimensional phenotype $\vec E(\vec\sigma)$: 

\EQ E^\alpha = \underbrace{ \sum_i J_i^\alpha \sigma_i }_{E^\alpha_{(1)}}+\underbrace{  \sum_{i\neq j}J_{ij}^\alpha \sigma_i \sigma_j }_{E^\alpha_{(2)}}+ \dots + E^\alpha_{(r_\text{max})}\label{gen-phen-map}
\EE 
where $\alpha = 1,\dots, k $ refers to the $\alpha$-coordinate of the $k$-dimensional phenotype $\vec E$ and the subscript  $r = 1,2,\dots, r_\text{max}$ is the degree of genetic interaction. Without a loss of generality we assume that genotypic loci are bi-allelic with $\sigma_i \in\{-1, 1\}$ indicating the state of locus $i$.

Mutations occur with a rate $\mu$ per site per generation, resulting in  a site flip: $-1\rightleftharpoons 1$. Thus, the change in the $\alpha$ component of the phenotype due to mutations follows:
 \EQA
 \nonumber  \Delta E^\alpha &=& \sum_{r=1}^{r_\text{max}} \Delta E^\alpha_{(r)}\\
\nonumber &=& -2 \mu \sum_{i_1} J_{i_1}^\alpha \sigma_{i_1} - 4 \mu  \sum_{i_1,i_2}J_{i_1i_2}^\alpha \sigma_{i_1} \sigma_{i_2}+\dots  -2r\mu \sum_{i_1,\dots,i_{r_\text{max}}} J_{i_1\dots i_{r_\text{max}}}^\alpha \sigma_{i_1}  \dots \sigma_{i_\text{max}}+ \mathcal{O}(\mu^2)  \\
 &=&  -2 \mu \sum_r r   E^\alpha_{(r)} + \mathcal{O}(\mu^2) 
 \EEA
 where we assumed that the mutation rate per locus per generation is small such that we can neglect terms of order $\mu^2$ and higher. Similarly, the change in the mean $\alpha$-phenotype $\x^\alpha = \overline{E^\alpha} $ due to mutations upto $\mathcal{O}(\mu^2)$  follows, $\Delta \x^\alpha = -2 \mu \sum_r r \x^\alpha_{(r)} $, where  $\x_{(r)}^\alpha = \overline{E_{(r)}^\alpha}$ is the intra-population mean of the $\alpha$-component of the phenotype. In this case, we can define a mutation potential 
 \EQ M = -  K^{-1} \mu \sum_{\alpha=1}^k \left( \sum_{r=1}^{r_\text{max}} r \x^\alpha_{(r)}\right)^2\EE
such that the change in phenotype due to mutations follows,  
\EQ  
 d\x = K \cdot \nabla M\EE
 
If the magnitude of mutation rate per locus per generation $\mu$ is small such that double mutations are significantly less likely than single mutations, a mutational potential function can be defined for phenotypic evolution. Dominance of double (or higher order) mutations results in mutational curls~\cite{Baxter:2007is,Mustonen:2010iga} and thus, a break down of  potential approximation.

Most of our analyses are applicable to general fitness and mutation landscapes. However, we characterize in detail the features of artificial selection to direct evolution on quadratic fitness and mutation landscapes, where phenotypes evolve by natural selection towards an evolutionary optimum~\cite{Fisher:1930wy}. In this case, the impacts of selection and mutation  follow linear functions in the high-dimensional phenotypic space, $\nabla F = - 2 C_0 \cdot  \x$, $\nabla M= - 2 L \cdot  \x$, where $\x$ denotes the shifted phenotype vector centered around its stationary state and $C_0$ and $L$ are selection and mutation matrices, respectively--- $L$ is a generalization of per-locus mutation rate in high dimensions. The evolutionary model in this case assumes a linear genotype-phenotype map (i.e., $r_\text{max} =1 $ in eq.~\ref{gen-phen-map}) and a non-linear quadratic phenotype-fitness map, which resembles biophysical models of global epistasis~\cite{Mustonen:2005bg,Kinney:2008tb,Manhart:2015eg,Nourmohammad:2017is,Otwinowski:2018jj,Otwinowski:2018jb,Husain:2020js}. We can formulate the evolution of mean phenotypes by,
\EQ
d\x = -2K  C\, \x\, \d t + \Sigma \, \d\W 
\label{eq.NatEvolQuad}\EE
where  $C\equiv N(C_0 +  K^{-1} \, L)$ is the effective adaptive pressure, scaled by the population size, which quantifies  the potential of a phenotype to  accumulate mutations under selection. The adaptive potential could in principle be measured directly using lineage tracking evolution experiments, in which impacts of a large number of adaptive mutations can be simultaneously probed~\cite{Blundell:2014if}.

In this work, we will use $F$ as a short hand for the adaptive landscape under natural selection, whose gradient characterizes the adaptive pressure, $\nabla F =-2C \x $ in eq.~\ref{eq.NatEvolQuad}. We have also rescaled  time with the effective population size (i.e., $t \to N t$), which is the coalescence time in neutrality~\cite{Kingman:1982bk}.

Similar to the mean, the covariance matrix $K$ is  a time-dependent variable, impacted by evolution. However,  fluctuations of covariance are much faster compared to the mean phenotype, and therefore, covariance can be  approximated by its stationary ensemble-averaged estimate~\cite{Nourmohammad:2013ty,NourMohammad:2016hg}. Moreover, even in the presence of moderately strong selection pressure, the phenotypic covariance depends  only weakly on the strength of selection  and is  primarily determined by the supply of mutations in a population~\cite{Nourmohammad:2013ty,Held:2014di}. Therefore, we also assume that the phenotypic covariance matrix remains approximately constant over time, throughout evolution. With these approximations,  evolution of  the mean phenotype can be described as a  stochastic process with  a constant adaptive pressure that approaches its stationary state over a characteristic equilibration time $\sim(2KC)^{-1}$.

The  stochastic evolution of the mean phenotype in eq.~\ref{eq.NatEvolQuad} defines an ensemble of evolutionary trajectories. We can characterize the statistics of these evolutionary paths by the dynamics of the underlying conditional probability density $P(\x',t'|\x,t)$ for a population to have  a mean phenotype $\x'$ at time $t'$, given its state $\x$ at an earlier time  $t<t'$.  The dynamics of this probability density  follows a high-dimensional Fokker-Planck equation~\cite{Nourmohammad:2013ty},

{\small
\EQA
\nonumber\frac{\partial}{\partial t} P(\x',t'|\x,t) =
 \left[ \frac{1}{2 N} \Tr K \nabla_{\x\x}  - \nabla( K \cdot \nabla F)   \right]P(\x',t'|\x,t)\\
 \label{eq:P_dynamics}
\EEA}
\hspace{-1ex}where we  introduced the compact notation, $\Tr K  \nabla_{\x\x}  \equiv \sum_{ij}K_{ij} \frac{\partial}{\partial x_i}\frac{\partial}{\partial x_j}$. As a result, the conditional  distribution of phenotypes follows an Ornstein-Uhlenbeck process, described by a  time-dependent multi-variate Gaussian distribution.

}

\section{Hamilton-Jacobi-Bellman equation for optimal  control}
\label{AppA}
We define a general stochastic evolutionary process for a population of size $N$ with an evolutionary drive due to natural selection and mutations $A(\x,t)$ and an external artificial selection $\u(\x,t)$, 
\EQ
d\x =( A(\x)  + \u(\x,t) )\d t +\Sigma(\x) \d\W
\label{eq:stochEvoU}
\EE
Here, time $t$ is measured in units of the coalescence time $N$ (i.e., the effective population size). $\d\W$ is a differential random walk due to an underlying Wiener process with an amplitude $\Sigma$, which is the square root of the phenotypic covariance matrix $K$:  $\Sigma^\top \Sigma\equiv K$. 
The stochastic evolution in eq.~\ref{eq:stochEvoU} defines an ensemble of phenotypic trajectories, the statistics of which can be characterized by a conditional probability density $P(\x,t|\x',t')$ for a population to have a phenotype $\x$ at time $t$, given its state $\x'$ at a previous time $t' <t$.  
 For a given artificial selection protocol $\u(\x,t)$, the conditional probability density evolves according to a Fokker-Planck equation~\cite{Risken:1966bi}, 
\EQ
\frac{\partial}{\partial t} P(\x,t|\x',t')=\left[\frac{1}{2} \Tr K \nabla_{\x\x}  - \nabla_\x \cdot \left(A(\x)  + \u(\x,t)\right)\right]P(\x,t|\x',t') \label{eq:FPEvoU}
  \EE
where we have used the short hand notation,  $\nabla_\x  \cdot \mathcal{O}=\sum_i  \frac{\partial}{\partial x_i } \mathcal{O}$ and $\Tr  K  \nabla_{\x\x}  \mathcal{O} = \sum_{i,j}K_{ij}\frac{\partial^2}{\partial x_i \partial x_j}  \mathcal{O}$,  as operators that act on the function $ \mathcal{O}$ in front of them.\\

The purpose of  artificial selection is to minimize a  cost function, 
{ \EQ
\Omega(\x,\u,t) =  V( \x,t ) + \frac{1}{2} \u^\top B \u 
\label{eq:Omega}\EE}
where $V( \x,t )\equiv V(| \x_t -\x^*|)$ is the cost for deviating from the desired target $\x^*$ during evolution and  $B$ is the cost for intervening with natural evolution and applying  artificial selection  $\u\equiv \u(\x,t)$. 

We define  the cost-to-go function $J(\x, t)$  as the expected value for the cumulative cost from time $t$ to end of the process $t_f$, subject to the evolutionary dynamics and under an optimal control $\u^*_{t\to t_f}$,  
\EQ
J(\x,t)=\text{min}_{\u_{t\to t_f}} \left\langle  Q(\x,{t_f}) +\int_t^{t_f}  \Omega(\x_s,\u_s)  ds\right\rangle
\EE
Here,  $Q(\x,{t_f}) \equiv Q(|\x_{t_f}-\x^*|) $ is the cost of deviation from the target   at the end point $t_f$, which in general can be distinct from the path cost $V(\x_t)$. We can  formulate a recursive relation for the cost-to-go function $J(\x, t)$,
{\small
\EQA
\nonumber J(\x,t) &=& \text{min}_{\u_{t\to t_f}} \left\langle  Q(\x_{t_f})+\int_{t}^{t_f} \Omega(\x_s ,\u_s) ds  \right\rangle\\
\nonumber &=&\lim_{\delta t\to 0} \text{min}_{\u_{t\to t_f} }\left\langle  Q(\x_{t_f})+ \int_{t}^{t+\delta t} \Omega(\x_s ,\u_s) ds + \int_{t+\delta t}^{t_f} \Omega(\x_s ,\u_s) ds   \right\rangle\\
\nonumber&=&\lim_{\delta t\to 0}  \text{min}_{\u_{t\to t_f} } \left\langle J(\x_{t+\delta t},t+\delta t) + \int_{t}^{t+\delta t} \Omega(\x_{s} ,\u_{s}) ds  \right\rangle\\
\nonumber&=&J(\x_t,t) + \text{min}_{\u_{t\to t_f} } \left\langle \Omega(\x_s ,\u_s)\delta t +  
\left [\frac{\partial }{\partial t} J(\x_t,t) + \left(A(\x_t) +\u\right)^\top(\nabla J) + \frac{1}{2}\sum_{ij}K_{ij} \frac{\partial}{\partial x_i}\frac{\partial}{\partial x_j} J \right]\delta t \right\rangle
\\\label{HJB-der}
\EEA}
where we used Ito calculus to expand the cost-to-go function, $J(\x_{t+\delta t},t+\delta t)$; see e.g. ref.~\cite{Gardiner:2004tx}. By reordering the terms in eq.~\ref{HJB-der}, we arrive at the  Hamilton-Jacobi-Bellman (HJB) equation,
\EQA
\nonumber-\frac{\partial }{\partial t} J(\x,t) &=&\min_\u\left[ \Omega(\x_t ,\u_t)+ (A(\x_t)+\u)^\top \cdot \nabla J +  \frac{1}{2}\Tr K\nabla_{\x\x} J\right]\\
&=& \min_\u\left[ \frac{1}{2} \u^\top B \u +\u^\top\cdot \nabla J \right]+V(\x)+ A(\x_t)^\top \cdot \nabla J  +  \frac{1}{2}\Tr K \nabla_{\x\x} J \label{HJB-eq}
\EEA
The functional form for the optimal artificial selection $\u^*$ follows by minimizing the right hand side of  eq.~\ref{HJB-eq} with respect to $\u$, 
\EQ \u^* = -B^{-1}  \nabla J. \label{eq_optimal_policy}
\EE

Therefore, the  time- and phenotype-dependent solution for the cost-to-go function $J(\x,t)$ determines the optimal protocol for artificial selection $\u^*(\x,t)$. By substituting the form of the optimal control $\u^*$ in eq.~\ref{HJB-eq}, we arrive at a non-linear partial differential equation for the cost-to-go function,
\EQA
-\frac{\partial }{\partial t} J(\x,t) &=&- \frac{1}{2}(\nabla J)^\top B^{-1}  \nabla J+ V(\x)+ A(\x_t)^\top \cdot \nabla J  +  \frac{1}{2}\Tr K\nabla_{\x\x} J
\EEA
which should be solved with a boundary condition  $J(\x,t_f) = Q(\x,{t_f})$ at the end point.   We introduce a new variable $\Psi = \exp[-{J}/\lambda]$ as the exponential of the cost-to-go function. The dynamics of $\Psi$ follows,
{\small  \EQA
\frac{\lambda}{\Psi}  \frac{\partial }{\partial t} \Psi 
&=&-\frac{\lambda^2}{2 \Psi^2}  (\nabla \Psi)^\top  B^{-1} \nabla \Psi +V(\x)- \frac{\lambda}{\Psi}A(\x_t)^\top \cdot (\nabla \Psi) -  \frac{\lambda}{2 } K \left[\frac{-1}{\Psi^2}( \nabla\Psi)^\top \cdot \nabla\Psi + \frac{1}{\Psi}\nabla_{\x\x}\Psi\right]
\EEA
}

The dynamics of $\Psi$ linearizes if and only if there exists a scalar $\lambda$ that relates the control cost to the covariance matrix such that   $B=\lambda K^{-1}$. This criteria is known as the path-integral control condition~\cite{Kappen:2005bn,Kappen:2005kb} by which we can map a generally non-linear control problem onto a linear stochastic process. The  path-integral control condition implies that the cost of artificial selection on each phenotype should be inversely proportional to the phenotype's fluctuations. In other words, artificially tweaking with highly conserved phenotypes should be more costly  than with variable phenotypes.  In this case, the HJB equation for the transformed  cost-to-go function $\Psi$  follows,
\EQ
 \frac{\partial }{\partial t} \Psi = - A(\x)^\top  \cdot \nabla \Psi - \frac{1}{2} \Tr K \nabla_{\x\x} \Psi + \frac{1}{\lambda} V(\x) \Psi \equiv - \L^\dagger \Psi
 \label{eq.psi}\EE
where $\L^\dagger$ is a linear operator acting on the function $\Psi$.  Equation~\ref{eq.psi}  has the form of a  backward Fokker-Planck equation  with the boundary condition at the end point $\Psi(\x,{t_f}) = \exp[-J(\x,t_f)/\lambda] = \exp[Q(\x,{t_f})/\lambda]$. We can define a conjugate function $P_u$ that evolves forward in time according to the  Hermitian conjugate of the operator $\L^\dagger$. This conjugate operator $\L$ can be characterized by  evaluating the inner product of the two functions,   
\EQA
\nonumber \left\langle \L P_u |\Psi\right\rangle =\left\langle P_u | \L^\dagger \Psi \right\rangle &=& \int d\x \,P_u (\x,t) \L^\dagger \Psi(\x,t) \\
\nonumber&=&\int d\x \, P_u (\x,t) \left( A(\x)^\top  \cdot \nabla \Psi + \frac{1}{2} \Tr K \nabla_{\x\x} \Psi - \frac{1}{\lambda} V(\x) \Psi\right) \Psi(\x,t)\\
\nonumber&=& \int d\x  \left( - \frac{1}{\lambda} V(\x) P_u (\x,t) - \nabla A(\x) P_u + \frac{1}{2}\Tr K \nabla_{\x\x} P_u\right)^\top \Psi (\x,t)\\\EEA
where we performed integration by part and assumed that the function $P_u$ vanishes at the boundaries. This formulation suggests a forward evolution by the operator $\L^\dagger$ for the function $P_u(\x',t'|\x,t)$ , 
\EQ
\frac{\partial}{\partial t} P_u  (\x',t'|\x,t) = \L P_u (\x',t'|\x,t) =  \left [ \underbrace{\frac{1}{2} \Tr K \nabla_{\x\x}  - \nabla A(\x)  }_{\L_0} - \frac{1}{\lambda} V(\x)\right]  P_u  (\x',t'|\x,t)
 \label{rho_dynamics}
\EE
with a boundary condition at the initial time point $P(\x',t|\x,t)= \delta (\x-\x')$. Importantly, the linear operator $\L$ resembles the forward Fokker Planck operator $\L_0$  for evolution under natural selection (i.e., the dynamics in eq.~\ref{eq:FPEvoU} with $\u=0$)
 with an extra annihilation term $V(\x)/\lambda $. The evolutionary dynamics with the $\L_0$ operator under natural selection conserves the probability density. The annihilation term on the other hand, eliminates the evolutionary trajectories  with a rate proportional to their cost  $V(\x,t)/\lambda$ at each time point.\\ 

Since $\Psi$ evolves backward in time according to $\L^\dagger$  and $P_u$ evolves forward in time according to $\L$, their inner product $\left\langle P_u | \Psi \right\rangle = \int d\x'P_u (\x',t'| \x,t)\Psi(\x',t')$ remains time-invariant\footnote{
The inner product of the two conjugate functions $\left \langle P_u | \Psi \right\rangle $ is time invariant:  

\EQA
\nonumber \left\langle P_u (t') | \Psi (t')\right\rangle \equiv \int d\x'P_u (\x',t'| \x,t)\Psi(\x',t')&=& \left\langle e^{\L (t'-t)} P_u(t) | e^{-\L^\dagger (t'-t)} \Psi(t) \right\rangle\\
\nonumber &=& \left\langle P_u(t) |  e^{\L^\dagger (t'-t)} e^{-\L^\dagger (t'-t)} \Psi(t) \right\rangle \equiv  \left\langle P_u(t) |\Psi(t) \right\rangle\EEA

}. 
Therefore, the inner product of the two functions at the initial and the final time points are equal, which follows,
\EQA
\nonumber \left\langle P_u (t) | \Psi (t)\right\rangle= \left\langle P_u (t_f) | \Psi (t_f)\right\rangle\rightarrow&& \int d\x' P_u (\x',t|x,t) \Psi(\x',t) =  \int d\x' P_u(\x',t_f|x,t) \Psi(\x',t_f)\\
 \rightarrow && \Psi (\x,t) = \int d\x'  P_u (\x',t_f| \x,t)\exp[-Q(\x',t_f)/\lambda]\EEA
 where we  substituted the  boundary condition for $P_u$ at the initial time  $t$ and for $\Psi$ at the finial time $t_f$. Thus, the cost-to-go function follows, 
\EQA
J(\x,t)= -\lambda \log \Psi(\x,t)&=& -\lambda \log \int d\x'  P_u (\x',t_f| \x,t)\exp[-Q(\x',t_f)/\lambda]
\label{eq.value-int}
\EEA

\section{Path integral solution to stochastic adaptive control}
\label{AppB}
Given the structure of the linear forward operator $\L$ (eq.~\ref{rho_dynamics}), we can either exactly compute the conditional function $P_u (\x',t_f| \x,t)$ or to use approximation methods common for  path integrals (e.g. the semi-classical method) to evaluate cost-to-go function in eq.~\ref{eq.value-int}. 
To formulate a path integral  for $P_u (\x',t_f| \x,t)$, we discretize the time window $[t:t_f]$ into $n$ small time slices of length $\epsilon$, ($t_0,\,t_1,\dots,\,t_n$), with $n\epsilon = t_f-t$. The conditional probability $P_u(\x',t_f | \x,t)$ can be written as an integral over all trajectories that start at the phenotypic state $\x$ at time $t_0\equiv t$ and end at $\x'$ at time $t_n\equiv t_f$,
\EQ
\label{eq.pathintegral}
P_u(\x',t_f |\x,t)\sim \int \prod_{i=1}^{n} d\x_i \,P_u(\x_i,t_i| \x_{i-1},t_{i-1} )\, \delta (\x_n - \x')\EE

The short-time  propagator $P_u(\x_i,t_{i}| \x_{i-1},t_{i-1} )$ follows a simple Gaussian form~\cite{Risken:1966bi}, 

\EQA
\nonumber P_u(\x_i,t_{i}| \x_{i-1},t_{i-1} ) &\sim& \exp\left\{-\frac{1}{ \lambda}\left[\Big (\x_i - \x_{i-1} - \epsilon A(\x_i)\Big)^\top\frac{\lambda K^{-1}}{2\epsilon} \Big (\x_i - \x_{i-1} - \epsilon A(\x_i)\Big) +V(\x_i )\epsilon\right]\right\}\\
&=&\exp\left\{-\frac{\epsilon}{\lambda} \left[\left(\frac{\x_i - \x_{i-1}}{\epsilon} -  A(\x_i)\right)^\top \frac{B}{2} \left (\frac{\x_i - \x_{i-1}}{\epsilon}  -  A(\x_i)\right) +V(\x_i )\right]\right\}
\EEA
where we used, $K = \lambda B^{-1}$. We can express the cost-to-go function (eq.~\ref{eq.value-int}) as a path integral,

{\small \EQA
\nonumber e^{-J(\x,t)/\lambda} &=& \int d\x'  P_u (\x',t_f| \x,t)\exp[-Q(\x',t_f)/\lambda]
\\
\nonumber&\sim& \int \prod_{i=1}^n d\x_i \exp\left\{-\frac{\epsilon}{\lambda} \left[\left(\frac{\x_i - \x_{i-1}}{\epsilon} -  A(\x_i)\right)^\top \frac{B}{2} \left (\frac{\x_i - \x_{i-1}}{\epsilon}  -  A(\x_i)\right) + V(\x_i ) +\frac{ Q(\x_n)}{\epsilon}\right]\right\} \\
\nonumber&\sim& \int \mathcal{D} (\x)  \exp\left[-\frac{1}{\lambda}  \left ( Q(\x(t_f)) + \int_t^{t_f} \d t\, \left[  \left(\frac{d \x(t) }{\d t} - A(\x(t),t)\right)^\top \frac{B}{2} \left(\frac{d \x(t) }{\d t} - A(\x(t),t)\right) +V(\x,t)\right] \right)\right]\\
&\equiv&\int \mathcal{D}(\x) \exp \left [-\frac{1}{\lambda} S_{\text{path}} (\x(t\to t_f))\right]
\EEA}
where $ S_{\text{path}} (\x(t\to t_f))$ is a corresponding action and $\mathcal{D}(\x)\sim \prod_{i=1}^n d\x_i$ is the integral measure over all the trajectories that start at $\x_0 = \x (0)$. Numerically, this formulation provides a way to generate evolutionary trajectories under artificial selection as an exponentially  weighted  ensemble from  trajectories under natural selection~\cite{Kappen:2005bn,Kappen:2005kb}. Moreover, 
if $\lambda$ is small, the  integral is dominated by the trajectories that are close to  the most likely (classical) trajectory $\hat \x(t\to t_f)$, and the path integral can be approximated using the semi-classical method; see ref.~\cite{Kappen:2005bn}.

\section{Control of molecular phenotypes with quadratic cost}
\label{AppC}

In the case that the path cost is zero $V(\x)=0$,  the artificially and naturally selected trajectories become distinct only due to the contributions from the end-point cost at $t=t_f$. For the choice of a linear evolutionary force  $A(\x) = -2 K C \x$ and a quadratic end-point cost, $Q(\x,{t_f})= \frac{1}{2} (\x_{t_f}-\x^*)^\top \tilde G (\x_{t_f}-\x^*)$,   evolution  follows  an Ornstein-Uhlenbeck (OU) process and the solution to eq.~(\ref{rho_dynamics})  takes  a  Gaussian form (see e.g. ref.~\cite{Gardiner:2004tx}),
{\small \EQA
\nonumber P_u(\x,t) &=& \int d\x_{t_f}   P_0 (\x_{t_f},t_f | \x,t)  P_0 (\x,t) \exp\left[- Q(\x_{t_f})/\lambda\right] \\
&\sim& \int d\x_{t_f}   \exp\left[ \frac{-1}{2}\left(\x_{t_f}- \mu(\x,\tau)\right)^\top K^{-1}_\tau\left (\x_{t_f}- \mu(\x,\tau)\right)\right]P_0 (\x,t) \exp\left[- Q(\x_{t_f})/\lambda\right]
 \EEA}
where $P_0(\x,t)$ denotes the marginal phenotype distribution in the uncontrolled process   at time $t$ and  $P_0(\x_{t_f},t_f | \x,t)$ indicates the conditional probability density in the uncontrolled process, which is a Gaussian distribution  with a  time-dependent mean,
 \EQA
 \mu(\x,\tau) &=& \exp[-2 K C \tau] \x, \EEA
 and a covariance matrix,
 \EQA
  \label{eq.cov-t}K_\tau &= & \int_t^{t_f} \d t' \exp[ -2 K C (t_f-t') ] K  \exp[ -2 C K (t_f-t') ]. 
 \EEA
with  $\tau= t_f-t$.\\

In this case, the cost-to-go in  eq.~(\ref{eq.value-int}) can be evaluated by a Gaussian integral to marginalize over the end state $\x_{t_f}$,
{\small \EQA
\nonumber\exp[-J/\lambda] &\sim& \int d \x_{t_f} \exp\left[ \frac{-1}{2\lambda}(\x_{t_f}- \mu(\x,\tau))^\top \lambda K^{-1}_\tau(\x_{t_f}- \mu(\x,\tau))\right]  \exp\left[\frac{-1}{2\lambda}  (\x_{t_f}-\x^*)^\top \G (\x_{t_f}-\x^*)\right]\\
&\sim& \exp\left[\frac{1}{2\lambda}  (\mu(\x,\tau)-\x^*)^\top 
\left(\G \left[\lambda K^{-1}_\tau +\G\right]^{-1} \G -\G\right)( \mu(\x,\tau)-\x^*)\right]
\EEA }
resulting in an optimal artificial selection protocol, 
\EQA
\nonumber  \u^* =  -B^{-1}  \nabla J &=& -\frac{ K}{\lambda } \nabla J = - \frac{K}{\lambda}   \left[\nabla \mu(\x,\tau) \right]^\top \left[\G-\G \left[\lambda K^{-1}_\tau +\G\right]^{-1} \G \right] (\mu(\x,\tau)-\x^*)\\
&=& -\frac{ K}{\lambda}  \exp[-2  C K \tau] \G \left[I- \frac{K_\tau}{\lambda} \left[ I +\frac{K_\tau}{\lambda}\G\right]^{-1} \G \right]  ( e^{-2 K C \tau} \x-\x^*) \label{eq.SI_Full_u}
\EEA

{When the  goal is to drive the population towards a target by an end point $t_f$,  the effective fitness $\hat F(\x,t)$ remains close to the natural landscape for an extended period. As time approaches the end point, populations transition from evolving in their natural landscape $F(\x)$ to  the artificially induced fitness landscape $F_{\text{art.}}(\x,t_f)$;  see Figs.~\ref{Fig:4},~S1,~S2 for evolution in one and two dimensions. Moreover, towards the end point, the  fitness peak and the strength of selection approach their final values, determined by the target and the cost functions in eq.~\ref{eq:Omega}, in an exponentially rapid manner (Figs.~S1,~\ref{Fig:4},~S2).
}

As the time approaches the end point ($t\to t_f$ or $\tau\to 0$), optimal artificial selection acts as a linear attractive force (i.e., a stabilizing selection)
\EQA
 \u^*(\tau \to 0) = \frac{-1}{\lambda} K\G\cdot (\x-\x^*)
+\mathcal{O}(\tau)
\label{ustartau0}\EEA 
to maintain the population close to the phenotypic target, with an effective strength of  artificial stabilizing selection $\G/\lambda$.

{
\section{Model misspecification and alternative approaches to artificial selection}
\label{AppAltSel}
Devising an optimal control strategy relies on the knowledge of system dynamics, which in some cases, may not be available with high precision. In such settings, ad-hoc selection protocols, guided by optimal control, can be used to drive evolution towards a desired target. Here, we consider such alternative (ad-hoc) selection protocols and compare their performance with respect to the optimal artificial selection strategy in terms of total cost and performance variability for the 2D covariate phenotypes considered in Section C. 

\begin{itemize}
\item[{\bf A)}] {\bf Control without the knowledge of phenotype correlation}\\
Optimal control on multi-variate phenotypes should be devised by considering the phenotypic correlation, which  could lead to non-monotonic artificial selection strategies in the course of evolution (Fig.~\ref{Fig:4}C). Here, we can quantify the importance of phenotypic covariance on devising optimal artificial selection protocols. To do so, we assume that the covariance matrix is misspecified and that phenotypes evolve independently (i.e.,  assuming the covariance matrix $K$ is diagonal). The devised protocol in this case does not take into account the synergistic or antagonistic interactions between the phenotypes and their response to interventions. Therefore, as correlation between phenotypes increase, the cost function (i.e., $ Q (\x,t_f)+  \int_t^{t_f} ds\left( V( \x_s ) + \frac{1}{2} \u_s^\top B \tilde\u_s \right)$) with the misspecified control protocol $\tilde \u$ soars (Fig.~S2). It should be noted  that although phenotypic cost with the misspecified control protocol always exceeds the cost under optimal control, its impact could still be favorable over the uncontrolled system (Fig.~S2), in certain parameter regimes.  

\item[{\bf B)}] {\bf Ad-hoc proportional  artificial selection protocol on multivariate phenotypes} \\
We consider a na\"ive (and intuitive) approach to artificial selection. Specifically, for 2D covariate phenotypes, the strategy is to intervene with a proportional control (i.e., artificial selection in a quadratic landscape) as long as the phenotypes are outside a specified neighborhood (with range $\delta$) of their targets, and to relax intervention once phenotypes are close enough to their targets:
\begin{equation}
\mathbf{u} = 
\begin{cases}
{\bf u}_x = -\kappa_t (x_t-x^*), \; {\bf u}_y=0 & \text{if }  |x_t -x^*|>\delta \\
{\bf u}_x = 0, \; {\bf u}_y=-\kappa_t (y_t-y^*) & \text{if }  |x_t -x^*|<\delta,\; |y_t -y^*|>\delta\\
{\bf u}_x = 0, \; {\bf u}_y=0 & \text{otherwise }  \\
\end{cases} \label{eq_prop_strategy}
\end{equation}
where  $\kappa_t$ is the strength of artificial selection. In this strategy, artificial selection is preferentially applied to phenotype ``x" and once ``x" is close to enough to its target, artificial interventions would selection for phenotype  ``y". In a sense, this protocol presents an intuitive approach to a control problem  similar to the one considered in  Fig.~\ref{Fig:4}, in which  phenotype ``x" is driven towards a target far from its natural state, while keeping phenotype ``y" close to its natural state.

We implement two versions of eq.~\ref{eq_prop_strategy}: In the first version, we select $\kappa_t$ to be equal to a constant $\kappa>0$ over all times. We denote this version as  proportional control.  In the second version, we implement an exponentially increasing artificial selection strength as evolution approaches the end time $T$, i.e., $\kappa_t=\kappa \exp^{-\tau+1}$. We denote this protocol as proportional control with exponential weights. We note that the choice of exponentially increasing artificial selection strength mimics the increase in control gains of optimal artificial selection strategy as remaining time approaches, in accordance with eq.~\ref{eq.SI_Full_u}. 
 We tune the constants in both control strategies such that  the mean values of the ensemble of end point values of phenotype $x$ ($\langle x_T\rangle$) are approximately equal to the mean ensemble value obtained from the optimal artificial selection protocol, and close to the desired target $x^*$.

We observe that the focal phenotype $x_t$ at the end point is much more widely distributed under both versions of the proportional control protocol (eq.~\ref{eq_prop_strategy}) compared to the optimal protocol (Fig.~S5A). Similarly, both proportional control strategies obtain similar $y_T$ values near the target $y^*=0$, with a higher ensemble variability than the one obtained using the optimal artificial selection strategy (Fig.~S5A). The variability in end point values tend to be higher in proportional control with exponential weights compared to the rest of the protocols. We also note that the peak magnitude of the artificial selection is much smaller for the optimal control compared to the proportional control protocols---see Fig. S6 for a comparison of control magnitudes between the optimal strategy and the proportional control strategy with exponential weights. 

When we compare the costs of control, the proportional control strategy with exponential weights has an order of magnitude smaller cost than the proportional control with a constant weight (Fig.~S5B). This is expected because the earlier control actions incur control costs but their effects are overwritten by the uncontrolled selection dynamics. In addition, the proportional control with exponential weights is on average preferred over doing nothing in terms of the total cost. Still, the accumulated cost for the proportional control strategy with exponential weights is at least an order of magnitude higher than the costs incurred under the optimal  strategy (Fig.~S5B). 

In summary, while the ad-hoc control strategies can be tuned to drive evolution towards a desired phenotypic target on average, they would suffer from an increased variability across evolutionary realizations and an elevated cost of control. 

\end{itemize}
}

\section{Artificial selection with intermittent monitoring}
\label{AppD}
Here, we assume that  artificial selection is imposed in discrete steps with time interval  $\tau$. Similar to the continuous control, the cost function has two components: the cost of control at the end of each intervention and  the cumulative cost of deviation from the optimum throughout each interval. The stationary cost-to-go function follows, 
{\small \EQA
&&J(\x,t_m;\tau) 
=\min_{\u}  \lim_{M\to \infty} \frac{1}{(M-m)\tau} \left\langle \sum_{i=m}^M \u_i^\top B \u_i + \int_{t_i}^{t_M} V(\x_t)\d t \right\rangle_{\text{evol.}}
\label{eq:Stat.J}
\EEA}
where we have normalized the path cost by the interval $\tau$ to assure that the cost-to-go remains finite.

To further simplify, we  only consider one dimensional phenotype $x$ with intra-population variance $k$, the cost of   deviation $V(x) = g(x-x^*)^2/2$ from target $x^*$, and the cost of intervention $\beta u^2/2$  with  artificial selection $u$. In the stationary state and in the regime of small interventions $(gk/\lambda<1)$, we assume that the optimal artificial selection protocol $u^*$ should be a variant of the case with full information with a strength of selection $\alpha_\tau$ dependent on the time window $\tau$, $u^*_\tau=-k \alpha_\tau (x-x^*)$. Our goal is to characterize the strength of optimal artificial selection $\alpha_\tau$.
 
The total cost over an interval $\tau$ in the stationary state follows,
 \EQA
 \Omega_\tau(x) &=&\frac{\beta}{2} \langle u^2\rangle + \frac{1}{\tau}  \left\langle \int_{t=t_i}^{t_i+\tau} V(x_t) \d t\right\rangle= \left \langle \frac{\beta }{2}k^2\alpha_\tau^2 (x_{\tau} - x^*)^2+ \frac{1}{2\tau} \gamma \int_{t=0}^\tau (x_t-x^*)^2 \d t\right\rangle 
 \label{eq.cumCostInter}
 \EEA

We are interested in the regime of moderate to weak interventions $(gk/\lambda<1)$, for which the linear response theory can characterize the  change in the state of the system after each intervention. In this regime,  evolution under artificial selection can be approximated as a perturbation from the dynamics under natural selection. The evolutionary dynamics of the  phenotype distribution is governed by a time-dependent Fokker Planck  operator, $L(x,t)$,
\EQ
\frac{\partial}{\partial t} P_{u;\tau}(x,t) = [L_{0}(\x) + L_{u}(\x) Y(t)] P_{u;\tau}(x,t)
\label{eq:FP_controlIntemittent}\EE
where $P_{u;\tau}(\x,t) $ is the full probability density under intermittent artificial selection, which can be split into the stationary solution  under natural selection  and the time-dependent deviation due to artificial selection: $P_{u;\tau}(x,t) = P_0(\x) + P_u(\x,t;\tau)$. 
 $L_{0} (\x) $ is the Fokker Planck operator associated with evolution under natural selection  (i.e., the dynamics in eq.~\ref{eq:FPEvoU} with $\u=0$),  $L_{u} (\x)= \partial_x k \alpha_\tau\,(x-x^*)$ is the state-dependent operator associated with artificial selection and $Y(t) =
 \lim_{M\to \infty} \sum_{i=1}^M \delta(t-t_i)
 $ characterizes a time-dependent component due to  artificial selection interventions at the end of each time interval. 

In the regime of linear response, where the impact of artificial selection is small, the deviation $\langle \Delta z\rangle$ of an expected value of an arbitrary function $\langle z(\x)\rangle$  from it stationary state (i.e., under natural selection) follows (see e.g. ref.~\cite{Risken:1966bi}), 
\EQ 
\langle \Delta z(t) \rangle = \int z(\x) P_{u}(x,t)  d\x \equiv \int_{-\infty}^{\infty} R_{z,L_u}(t-t') Y(t') \d t'
  \EE
where $R_{z,L_u} (t)$ is the response function to artificial selection $L_u$,
\EQA
R_{z,L_u} (t) &=&\begin{cases}
 \int z(\x) \,e^{L_{0} (\x) \,\cdot t} \,\,L_{u} (\x)  P_0(\x) d\x & \text{for } t\geq 0\\\\
 0 &  \text{for } t< 0\\
 \end{cases}
 \EEA

At end of each time interval,  artificial selection imposes a shock-type perturbation to the evolutionary process. The immediate response of the population to this selection pressure can be characterized by the instantaneous response function (i.e., with $Y(t')=\delta (t-t')$), resulting in  the change in a given phenotypic statistics $z$ (see e.g. ref.~\cite{Risken:1966bi}),
\EQA
\nonumber \langle \Delta z(t) \rangle&=& \int z(x) \L_u(\x) P_0(x) dx\\
\nonumber&=& \frac{1}{Z} \int dx\, z(x) \,\frac{\partial }{\partial x} \left(k\alpha_\tau\, (x- x^*)\, \exp\left[ -\frac{ x^2 } { 2\var_{\st}}\right] \right) \\
\nonumber&=&k \alpha_\tau \left( \langle z(x)\rangle_{\st} - \frac{1}{\var_{\st}}\left[  \left \langle z(x) \, x^2\right\rangle_{\st}  -x^* \langle z(x) \,x\rangle_{\st}  \right] \right)\\
 \EEA
where $P_0(x)$  is the Gaussian stationary distribution for the mean phenotype under natural selection,  $\var_\st= 1/4c$ is the stationary ensemble variance for the mean phenotype under natural selection, $Z$ is the normalization factor for the underlying stationary distribution, and
$\langle \cdot\rangle_\st$ indicates expectation values under the stationary distribution.

 At the beginning of each interval $t=0$ the deviation of the mean phenotype from its expectation under natural selection $\langle x\rangle_\st=0$ follows, 
\EQA
\langle \Delta x \rangle = \langle x (t=0)\rangle=k \alpha_\tau x^*\EEA

Similarly, the deviation in the second moment of the phenotypic distribution from the expectation under natural selection $\langle x^2\rangle_\st = \var_\st$  follows, 
\EQA
\langle \Delta x^2 \rangle = \langle x^2 (t=0)\rangle - \var_{\st} = \alpha_\tau  \left(\var_{\st} - \frac{1}{\var_{\st}} \langle x^4\rangle_{\st}\right) = - 2 k \alpha_\tau \,\var_{\st}= -k\alpha_\tau/2 c\EEA

Thus,   the phenotypic variance at the beginning of each interval follows,
\EQA 
\var_u (t=0)= \langle x^2 (t=0)\rangle -  \langle x (t=0)\rangle^2 = \var_{\st}\left[1- 2k\alpha_\tau\right] - \left[ k\alpha_\tau\, x^*\right]^2\EEA

Following an intervention at time $t=0$, populations evolve according to natural selection until the next intervention. Therefore, the phenotype statistics during each  time interval ($0<t<\tau$ ) follow, 
{\small \EQA 
\label{eq:TimeExp} \langle x(t)\rangle &=&  \alpha_\tau  x^* e^{-2k c t}\\\nonumber \\
\nonumber\var(t) &= &\left[ \var_{\st}(1- 4k c) - \left( k\alpha_\tau\, x^*\right)^2\right] e^{-4k c t} +  \label{eq:TimeVar}\var_{\st} (1-  e^{- 4k c t}) = \var_{\st} (1-2k\alpha_\tau
\, e^{-4k c t}) - \left( k\alpha_\tau\, x^*\right)^2e^{-4k c t}\\
 \EEA}
  
 We can now evaluate the cumulative cost function (eq.~\ref{eq.cumCostInter})
 \EQA
\nonumber \Omega_\tau(x)&=& \frac{\beta}{2} \langle u^2\rangle + \frac{1}{2\tau} \gamma \left\langle \int_{t=0}^\tau (x_t-x^*)^2 \d t\right\rangle\\
\nonumber\nonumber&=& \left \langle \frac{\beta }{2} k^2 \alpha_\tau^2 (x_{\tau} - x^*)^2+ \frac{1}{2\tau} \gamma \int_{t=0}^\tau (x_t-x^*)^2 \d t\right\rangle\\
\nonumber&=& \frac{\beta }{2}k^2\alpha_\tau^2 \left[ ( \langle x_{\tau}\rangle  - x^*)^2 +\var(\tau) \right ]+ \frac{1}{2\tau} \gamma \int_{t=0}^\tau\left[ (\langle x_t \rangle -x^*)^2  + \sigma^2(t)\right] \d t \\
\nonumber &=&\frac{\beta }{2} k^2\alpha_\tau^2\left[ \var_{\st}(1-2 k\alpha_\tau e^{-4k c \tau} ) + (x^*)^2 (1- 2k \alpha_\tau e^{-2k c \tau})\right]\\ 
 &&+ \frac{1}{2\tau} \gamma\left[\frac{-\alpha_\tau}{2 c}  \left((1-e^{-4k c \tau}) \var_{\st} + 2 (1- e^{-2k c \tau}) (x^*)^2\right) + (\var_{\st}+(x^*)^2)\tau
 \right]\label{eq:IntermittCost}
\EEA
where we have used the time-dependent expectation and variance in eqs.~\ref{eq:TimeExp} and \ref{eq:TimeVar}.

The optimal strength of artificial selection $\alpha^*_\tau$ for intermittent interventions  can be characterized by minimizing the cost function (eq.~\ref{eq:IntermittCost}) with respect to $\alpha_\tau$,

\EQA
\alpha^*_\tau&= & \frac{\gamma}{\lambda} \left(\frac{(1- e^{-\tau}) (1+ 8c(x^*)^2 +e^{-\tau})}{2\tau (1+ 4c (x^*)^2)} \right) + \mathcal{O}\left( \left(k\gamma/\lambda \right)^2\right)
\label{eq:alpha}
\EEA
which in the limit of small separation time ($\tau\to 0$) approaches the expectation under continuous monitoring in the stationary state (eq.~\ref{ustartau0}), $\alpha^*(\tau\to 0) = \gamma/\lambda$.

\section{Work performed by artificial selection}
\label{AppE}

Artificial selection changes the distribution of  phenotypic trajectories $\x_{t_0}^{t_f} \equiv (\x_{t_0},\dots,\x_{t_f})$ from $P_0(\x_{t_0}^{t_f}) $ in the stationary state under natural selection to a configuration   closer to the desired target $P_\u(\x_{t_0}^{t_f})$. In analogy to thermodynamics, we can associate a free energy to these distributions, as $F_0= \log P_0(\x_{t_0}^{t_f})$ and $F_u=\log P_u(\x_{t_0}^{t_f})$~\cite{Landau:1958up}. The expected difference between the free energy of the two phenotypic configurations can be interpreted as the amount of work done by artificial selection, which corresponds to the Kullback-Leibler distance between the two distributions,
\EQ
W_u \equiv \langle F_u \rangle -\langle F_0\rangle= \int \D\x\, P_u(\x_{t_0}^{t_f} ) \log \left[ \frac{P_\u(\x_{t_0}^{t_f} ) }{P_0(\x_{t_0}^{t_f} )}\right] \equiv D_{KL} (P_\u(\x_{t_0}^{t_f} )| | P_0(\x_{t_0}^{t_f} ))\label{eq.work}
\EE
where $\D\x$ is the integral measure over all trajectories. The estimate of work in eq.~\ref{eq.work} however  should not be interpreted as a physical work, rather as an information theoretical measure of discrimination between the two phenotype distributions due to artificial selection.

The evolution of the distribution for phenotype trajectories $P_u(\x_{t_0}^{t_f})$ under a given artificial selection protocol, $\u_{t_0}^{t_f}$ is Markovian  (eqs.~\ref{eq:stochEvoU},\ref{eq:FPEvoU}). 
To characterize this path probability density, we will follow the path integral formulation in eq.~\ref{eq.pathintegral} and discretize the time window $[t_0:t_f]$ into $n$ small time slices of length $\epsilon$, ($t_0,\,t_1,\dots,\,t_n$), with $n\epsilon = t_f-t$. The  probability of a given trajectory $P_u(\x_{t_0}^{t_f})$ can be written as a product of short-term propagators (i.e., conditional probabilities); see  ref.~\cite{Risken:1966bi},
\EQA
\nonumber P_u(\x_{t_0}^{t_f}) &=& \lim_{\epsilon\to 0}  \prod_{s=t_0}^{t_f} P_u(\x_{s+\epsilon} |\x_{s}) )\\
\nonumber&=&  \lim_{n\to \infty}  \prod_{i=1}^{n} \frac{1}{Z_i} \exp\left[-(\x_{i+1} - \x_i  -\epsilon (A(\x_i) +\u(\x_i) ) )^\top \frac{K^{-1}}{2\epsilon} (\x_{i+1} - \x_i  -\epsilon (A(\x_i) +\u(\x_i) ) )\right]\\
\nonumber&\sim& P_0(\x_{t_0}^{t_f}) \lim_{n\to \infty}  \prod_{i=1}^{n} e^{ \u^\top(\x_i) K^{-1} (\x_{i+1} - \x_i  -\epsilon A(\x_i) )} \times e^{- \frac{\epsilon}{2} \u(\x_i)^\top {K^{-1}} \u(\x_i) } \\
\nonumber&=&P_0(\x_{t_0}^{t_f})\exp\left [-\int_{t_0}^{t_f}  \d t \frac{1}{2} \u^\top(\x,t) K^{-1} \u(\x,t)  +\int_{t_0}^{t_f}  \u^\top(\x,t)   K^{-1} (d\x_t -  A(\x_t) \d t) \right] \\ \EEA

The Kullback-Leibler distance between the two distributions follows,

{\small \EQA
\nonumber D_{KL} (P_\u(\x_{t_0}^{t_f}) || P_0(\x_{t_0}^{t_f})) &=& \Large\int \D\x \,P_\u(\x_{t_0}^{t_f})\,  \left [-\int_{t_0}^{t_f}  \d t \left(\frac{1}{2} \u^\top(\x,t) K^{-1} u(\x,t) \right) +\int_{t_0}^{t_f}  \u^\top(\x,t)   K^{-1} (d\x_t -  A(\x_t) \d t) \right]\\
 &=&\Large\int_{t_0}^{t_f} \d t \int  \D\x \,P_\u(\x_{t_0}^{t_f})  \left( \frac{1}{2} \u^\top(\x,t) K^{-1} \u(\x,t)   \right)\equiv \left \langle   \frac{1}{2} \left(\u_{t_0}^{t_f}\right)^\top  K^{-1} \u_{t_0}^{t_f} \right\rangle \EEA}
where we have used $ d\x_t -  A(\x_t) \d t = \u(\x_t,t) \d t+ \d\W_t$, with $\d\W_t$ as the stochastic differential measure for a multi-variate Wiener process (see  ref.~\cite{Gardiner:2004tx}). Importantly, with the criteria of path integral control  (i.e.,  $K^{-1} = B/\lambda$), the Kullback-Leibler distance between the artificially and naturally  selected phenotype  distributions is equivalent to the cumulative cost of intervention, divided by the overall cost of   artificial selection $\lambda$,
\EQ
D_{KL} (P_\u(\x_{t_0}^{t_f}) || P_0(\x_{t_0}^{t_f})) = \frac{1}{2\lambda} \left \langle\left(\u_{t_0}^{t_f}\right)^\top B\, \u_{t_0}^{t_f}\right\rangle \EE
which can  intuitively be  interpreted as  the amount of work done by artificial selection.


%

\newpage{}
\vspace{10ex}
\renewcommand{\thefigure}{S\arabic{figure}}

\begin{figure}[h!]
\centering\includegraphics[width =0.5\columnwidth]{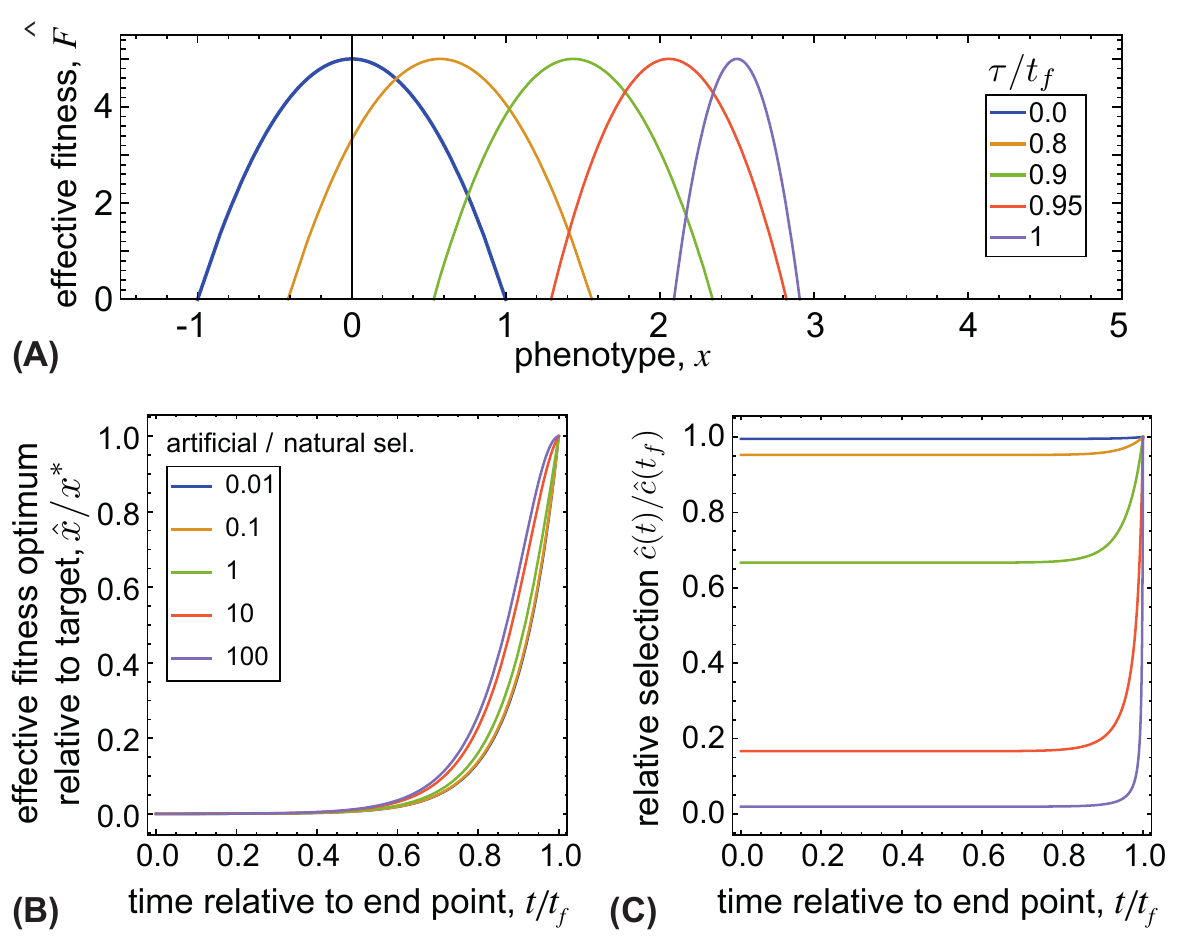}
\caption{{{\bf Optimal artificial selection for a 1D phenotype.} The impact of artificial selection intensifies as time approaches the end point of the process. {\bf(A)} The interplay between artificial and natural selection defines an effective  time-dependent (colors) fitness landscape $\hat F$ with an optimum $\hat x(t)$ that approaches the phenotypic target for artificial selection ($x^*=3$) and an effective selection pressure $\hat c$ that intensifies  as time approaches end point {$t/t_f\to 1$}.  Other parameters: $\lambda=0.1$; $c=2$; $g=2$. {\bf (B)} and {\bf (C)} show the  effective fitness peak relative to the target $\hat x/x^*$ and the relative selection pressure of the effective fitness landscape $\hat c(t)/\hat c(t_f)$ as a function of time, for a range of relative artificial to natural selection pressures $g/\lambda c$ (colors). }
\label{Fig:3} }
\end{figure}

\begin{figure}[h!]
\centering\includegraphics[width =\columnwidth]{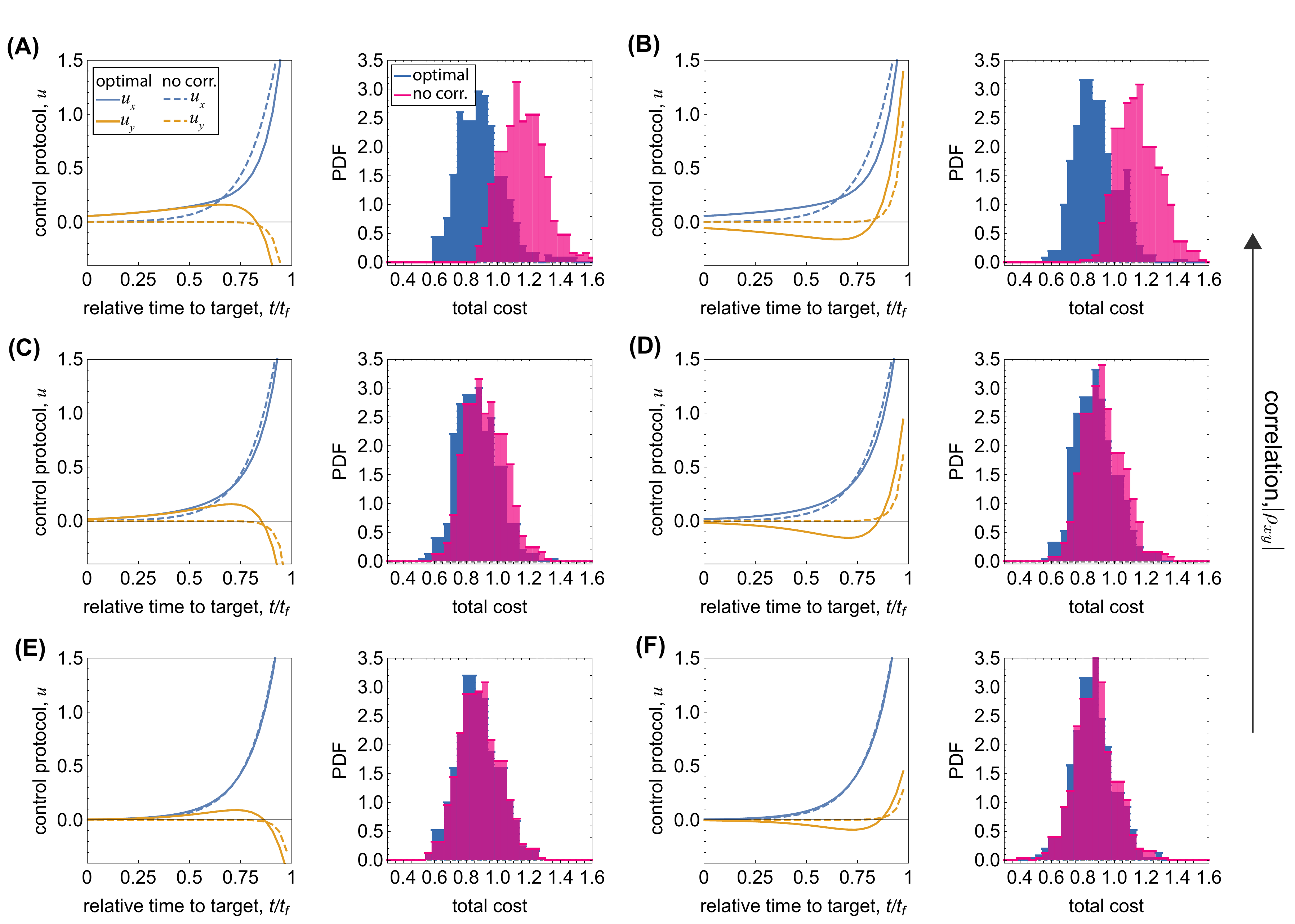}
\caption{
{{\bf Impact of covariance on artificial selection for multivariate phenotypes.} 
{\bf (A-F)} The left  panels show the optimal control protocol on 2D phenotypes (i) assuming the correct covariance matrix between the correlated phenotypes (full line), and by model misspecification, whereby the controller  assumes that phenotypes are uncorrelated (dashed line). The right panels show the distribution of the total control cost (for control with end-point cost), $Q (\x,t_f)+  \int_t^{t_f} ds \frac{1}{2} \u_s^\top B \u_s$ over 500 realizations of the evolutionary process, under optimal control (blue), under optimal control with misspecified information, assuming that  phenotypes are uncorrelated (pink), and without control (dashed line).  Parameters: $x^*=3$, $y^*=0$; $c_x=c_y =5$, $c_{xy}=0$;  $k_x=k_y=0.05$; $g_x=g_y=3$, $\lambda=0.01$, $t_f= 15 [N]$, and phenotypic correlations: (A) $\rho_{xy} = 0.75$, (B) $\rho_{xy} = -0.75$, (C) $\rho_{xy} = 0.5$, (D) $\rho_{xy} =- 0.5$, (E) $\rho_{xy} = 0.25$, (F)  $\rho_{xy} = -0.25$.}
\label{Fig:S3} }
\end{figure}

\begin{figure}[h!]
\includegraphics[width =0.85\columnwidth]{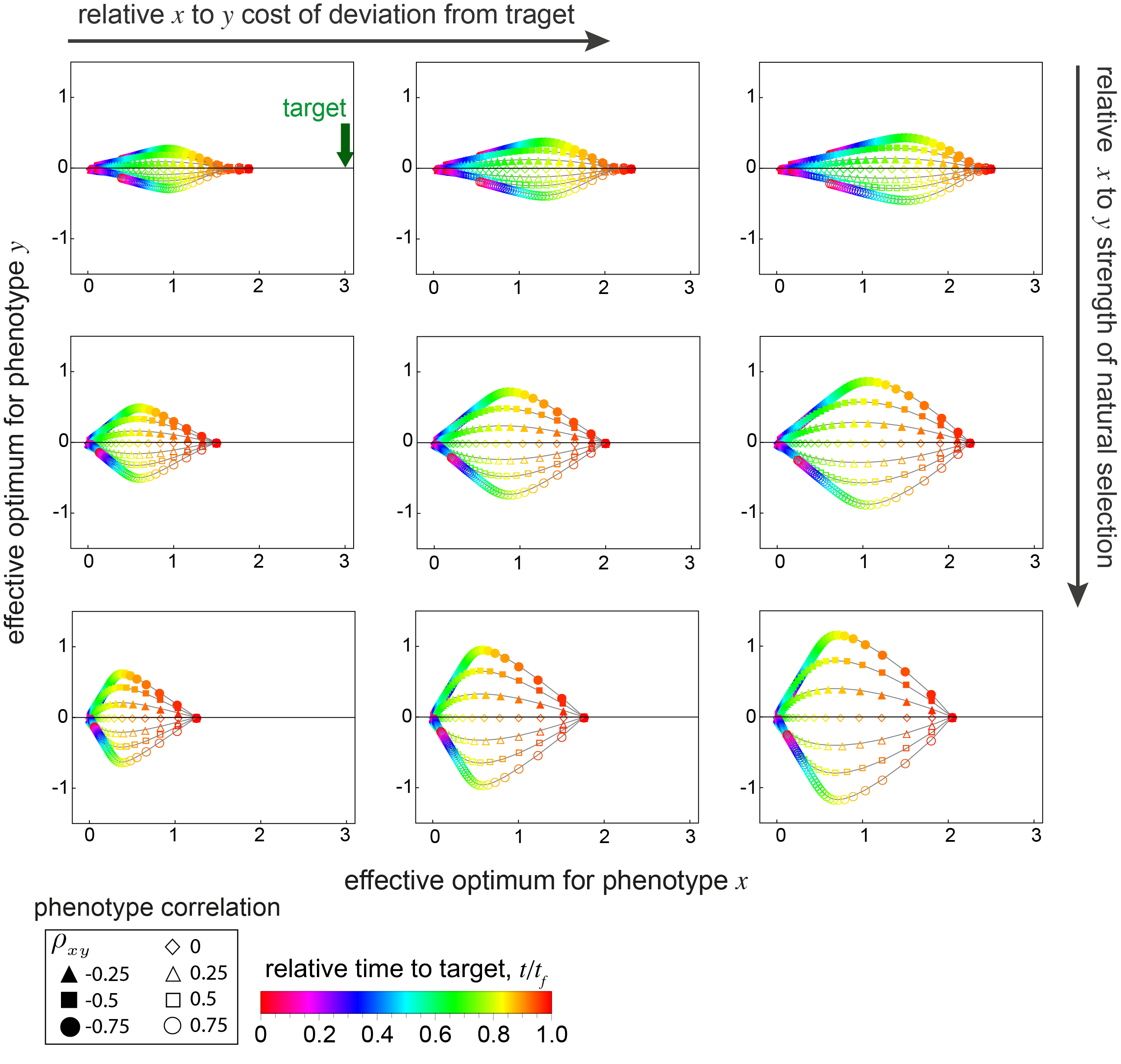}
\caption{{\bf Effective fitness optimum  for 2D covarying phenotypes under artificial selection.} 
The dynamics of the effective fitness peak  is shown over time (colors)  for 2D covarying phenotypes with correlations $\rho_{xy}$ indicated by the shape of the markers, and for increasing end-point cost of deviation from target along the $x$-phenotype,  $g_x= 1,\,2,\, 3$ from left to right, with $g_y=2$ and for increasing strength of natural selection on the $x$-phenotype, $c_x=3,\,5,\, 7$ from top to bottom with $c_y= 5$. Other parameters: $x^*=3$, $y^*=0$; $c_{xy}=0$;  $k_x=k_y=0.02$;  $\lambda=0.1$.
\label{Fig:S1} }
\end{figure}

\begin{figure}[h!]
\includegraphics[width =0.9\columnwidth]{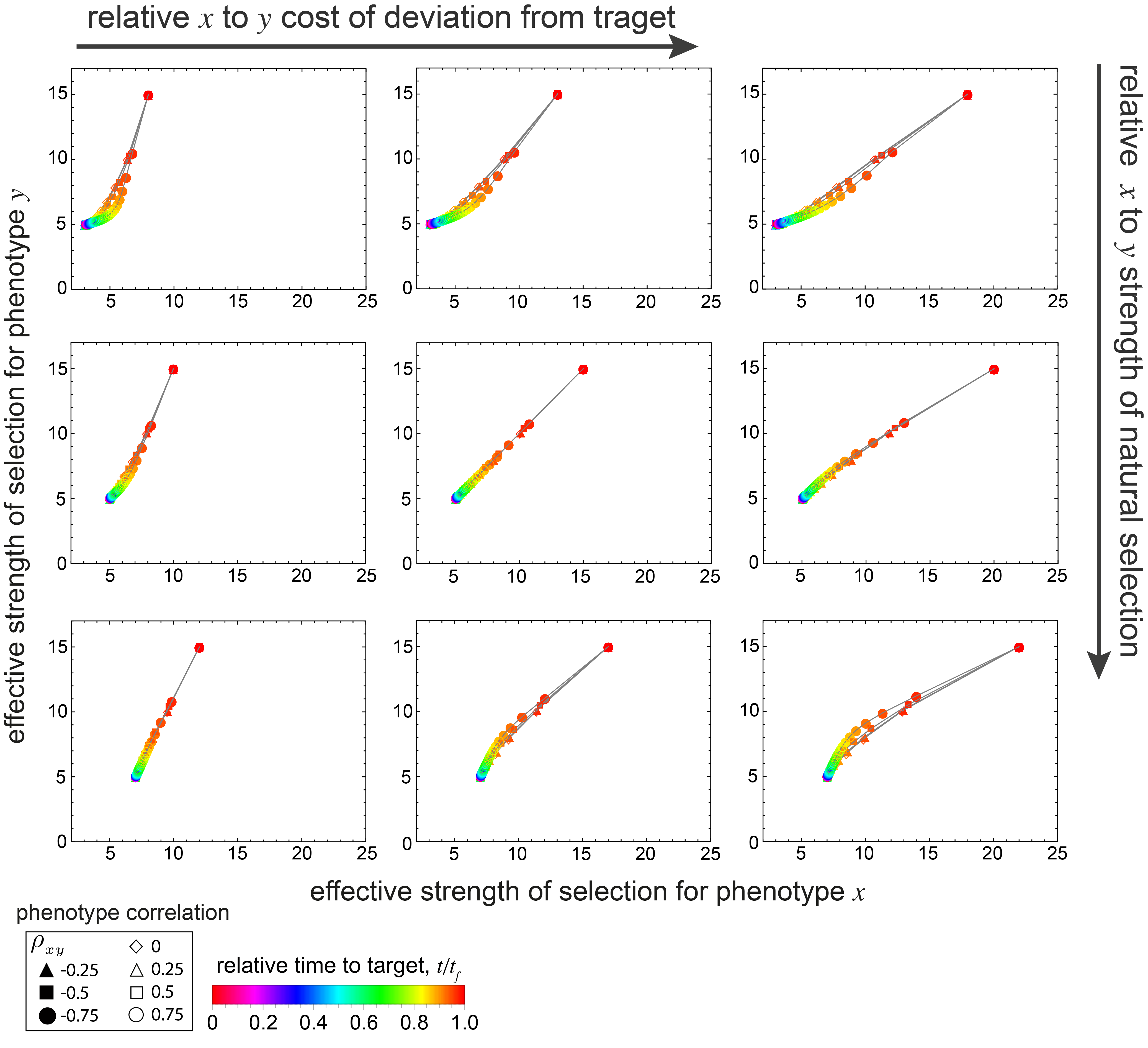}
\caption{{\bf Effective strength of selection for 2D covarying phenotypes  under artificial selection.} 
The dynamics of the effective strength of selection is shown over time (colors)  for 2D covarying phenotypes with correlations $\rho_{xy}$ indicated by the shape of the markers. The parameters in each panels are the same as in Fig.~\ref{Fig:S1}.
\label{Fig:S2} }
\end{figure}

\begin{figure}
\includegraphics[width=1\linewidth]{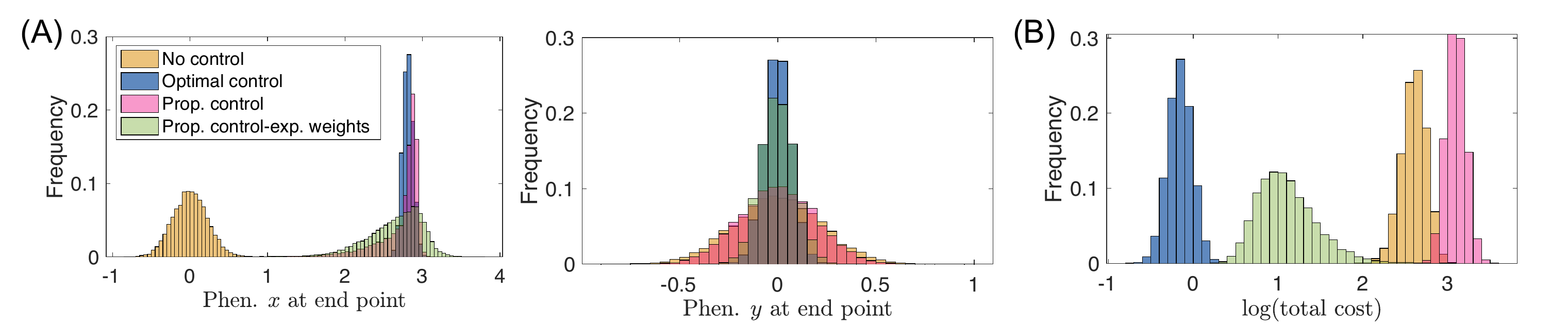}
\caption{{{\bf Optimal controller versus na\"{i}ve controllers.} {\bf (A)} The distribution of phenotypes at the end point of optimal artificial selection protocol compared to the phenotypic distribution under natural selection, proportional control, and proportional control with exponential weights. {\bf (B)} The distribution of logarithm of the total costs $\log(J(\mathbf{x}_0,0))$ for natural selection, optimal artificial selection protocol, proportional control, and proportional control with exponential weights. Evolutionary parameters are $c_x= 5$, $c_y=5$, $c_{xy}=0$; $x^*=3$, $y^*=0$; $k_x=0.05$; $k_y=0.05$; $k_{xy}=0$; $g_x=g_y=3$; $\lambda=0.01$. We let $\kappa_{x,t}=k_x g_x/\lambda=15$,  $\kappa_{y,t}= k_y g_y/\lambda=15$ and $\delta = 0.1$ in eq. (E1) for proportional control. Similarly, the exponential weights are set as  $\kappa_{x,t}= 2.5\exp(1-\tau)$,  $\kappa_{y,t}=2.5\exp(1-\tau)$ where $\tau$ denotes the remaining time. The proportional control gains are set so that the ensemble mean of the phenotype $x$ at the end point matches that of the optimal artificial selection protocol.}}\label{Fig:S5}
\end{figure}

\begin{figure}
\centering
\includegraphics[width=0.5\linewidth]{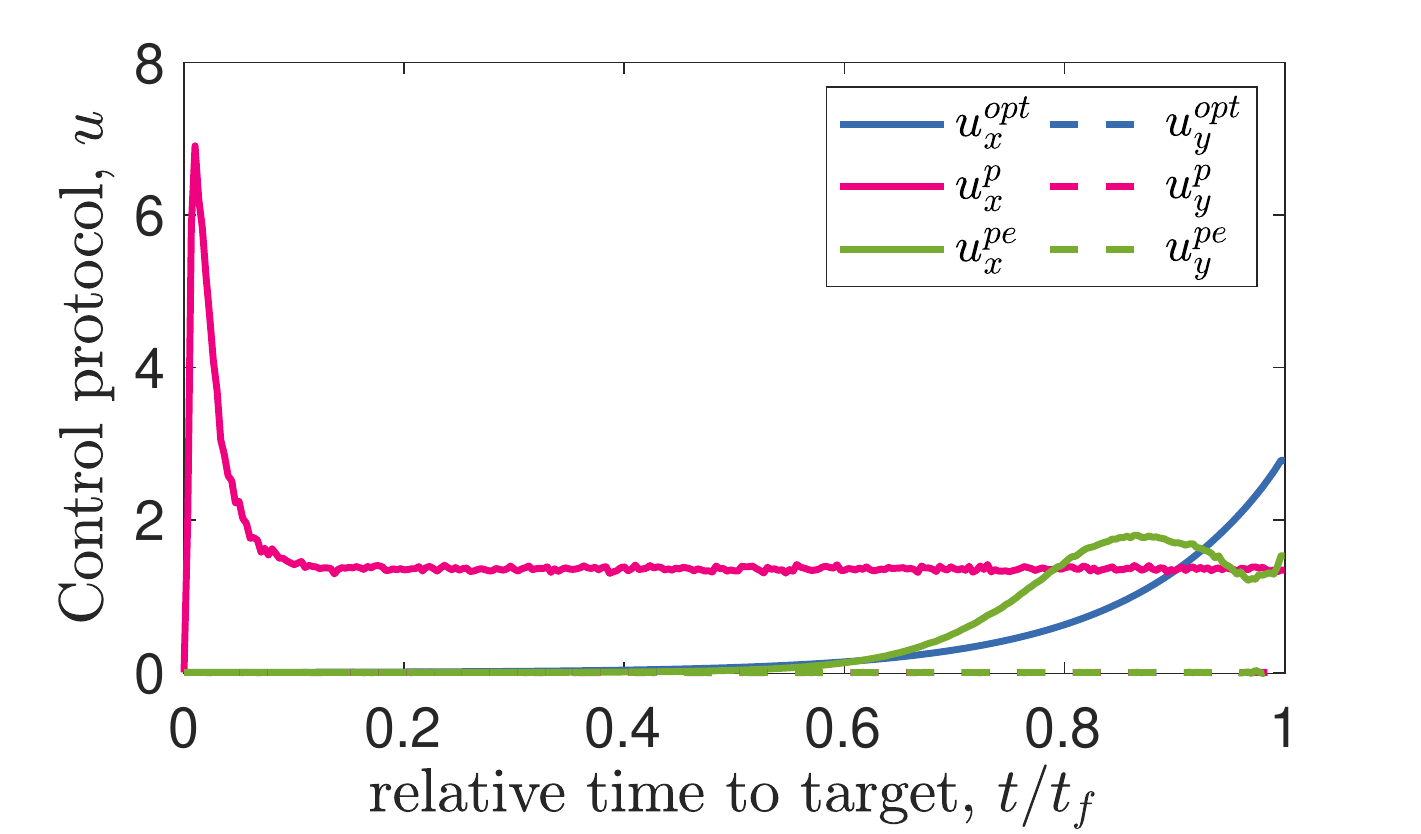}
\caption{{{\bf Alternative artificial selection protocols.} Ensemble mean of control actions for 2D phenotypes ($x,\,y$) under optimal artificial selection $u_x^{opt},\,u_y^{opt}$, proportional control $u_x^{p},\,u_y^{p}$, and proportional control with exponential weights $u_x^{pe},\,u_y^{pe}$ are shown as a function time to the target. Evolutionary and artificial selection parameters are similar to Fig. \ref{Fig:S5}.}}\label{Fig:S6}
\end{figure}

\end{document}